\documentclass[aps,prfluids,superscriptaddress,amsmath,amssymb,longbibliography]{revtex4-2}

\usepackage[utf8]{inputenc}
\usepackage{graphicx}
\usepackage{epstopdf, epsfig}
\usepackage{ulem}
\usepackage[dvipsnames]{xcolor}%

\newcommand{\bv}[1]{\mathbf{#1}} 
\newcommand{\dd}{\mathrm{d}} 

\newcommand{\activity}{\mathcal{A}}
\newcommand{\mobility}{\mathcal{M}}
\newcommand{\vslip}{\bv{v}_{\mathrm{slip}}}

\newcommand{\no}{\bv{n}_f}
\newcommand{\idmat}{\bv{1}}
\newcommand{\x}{\bv{x}}
\newcommand{\y}{\bv{y}}

\newcommand{\Greendiff}{\mathcal{G}}
\newcommand{\R}{\bv{R}}
\newcommand{\xo}{\tilde{\bv{x}}}
\newcommand{\length}{\ell}
\newcommand{\rc}{\bv{r}}
\newcommand{\surf}{\bv{S}}
\newcommand{\erho}{\hat{\bv{e}}_{\rho}}
\newcommand{\etheta}{\hat{\bv{e}}_{\theta}}
\newcommand{\tanhat}{\hat{\bv{t}}}
\newcommand{\norhat}{\hat{\bv{n}}}
\newcommand{\binorhat}{\hat{\bv{b}}}
\newcommand{\thetators}{\theta_i}
\newcommand{\crossradius}{\rho}
\newcommand{\maxcrossradius}{r_f}
\newcommand{\torsion}{\tau}
\newcommand{\curvature}{\kappa}
\newcommand{\dds}[1]{\frac{\partial {#1}}{\partial s}}
\newcommand{\ddth}[1]{\frac{\partial {#1}}{\partial \theta}}
\newcommand{\ddsprime}[1]{\frac{\partial {#1}}{\partial \sdum}}
\newcommand{\ddthprime}[1]{\frac{\partial {#1}}{\partial \thetadum}}

\newcommand{\dtheta}{\Theta}

\newcommand{\bvR}{\bv{R}}
\newcommand{\bvRo}{\bv{R}_0}
\newcommand{\epsslend}{\epsilon}  

\newcommand{\bvRione}{\hat{\bv{R}}^{(1)}_{(i)}}
\newcommand{\bvRitwo}{\hat{\bv{R}}^{(2)}_{(i)}}

\newcommand{\kerone}{K_{1}}
\newcommand{\kertwo}{K_{2}}
\newcommand{\outerexp}[1]{{#1}^{(\text{o})}}
\newcommand{\innerexp}[1]{{#1}^{(\text{i})}}
\newcommand{\expouterkernelininner}[1]{{#1}^{(\text{o})\in(\text{i})}}
\newcommand{\expinnerkernelinouter}[1]{{#1}^{(\text{i})\in(\text{o})}}
\newcommand{\sign}[1]{\mathrm{sign}\left({#1}\right)}

\newcommand{\logfactor}{L}

\newcommand{\lambdaf}{\lambda}

\newcommand{\zerothorder}[1]{{#1}^{(0)}}
\newcommand{\firstorder}[1]{{#1}^{(1)}}

\newcommand{\thetamode}{\Theta(s,\theta)}

\newcommand{\bvUo}{\bv{U}_0}
\newcommand{\bvfo}{\bv{f}_0}
\newcommand{\Uswim}{\bv{U}_{\mathrm{sw}}}
\newcommand{\Omegaswim}{\boldsymbol{\Omega}_{\mathrm{sw}}}

\newcommand{\ddrhods}{\frac{\dd \crossradius(s)}{\dd s}}

\newcommand{\sdum}{\tilde{s}}
\newcommand{\thetadum}{\tilde{\theta}}

\newcommand{\ccoeffcosmode}[1]{c_{c,{#1}}}
\newcommand{\ccoeffsinmode}[1]{c_{s,{#1}}}
\newcommand{\Acoeffcosmode}[1]{\activity_{c,{#1}}}
\newcommand{\Acoeffsinmode}[1]{\activity_{s,{#1}}}
\newcommand{\Ao}{\activity_0}

\newcommand{\Deltaerhorhoq}{\boldsymbol{\mathcal{D}}^{(s,\theta)}_{(s+q,\thetadum)}}
\newcommand{\asinh}[1]{\text{asinh}{#1}}

\newcommand{\czeroordercoeffcosmode}[1]{c^{(0)}_{c,{#1}}}
\newcommand{\czeroordercoeffsinmode}[1]{c^{(0)}_{s,{#1}}}
\newcommand{\czeroorderzeromode}{c^{(0)}_{{0}}}

\newcommand{\stresslet}{\underline{\underline{\bv{S}}}}
\newcommand{\stress}{\underline{\underline{\boldsymbol{\sigma}}}}
\newcommand{\Gammatensor}{\underline{\underline{\boldsymbol{\Gamma}}}}

\newcommand{\vslipzeromode}{\bv{v}^{\mathrm{slip}}_{0}}
\newcommand{\vslipcoeffcosmode}[1]{\bv{v}^{\mathrm{slip}}_{c,{#1}}}
\newcommand{\vslipcoeffsinmode}[1]{\bv{v}^{\mathrm{slip}}_{s,{#1}}}

\newcommand{\sinfunction}{\sigma}
\newcommand{\tanfunction}{\tau}

\newcommand{\deformcentrelineamp}{b}
\newcommand{\deformcentrelineshape}{n}
\newcommand{\deformradiusamp}{a}
\newcommand{\deformradiusnum}{k}

\usepackage{tikz,pgfplots}
\usetikzlibrary{fadings}
\usetikzlibrary{decorations.pathmorphing}
\usetikzlibrary{decorations.markings}
\tikzset{snake it/.style={decorate, decoration=snake}}

\usepackage[caption=false]{subfig} 

\begin{document}

\title{Slender Phoretic Loops and Knots} 

\affiliation{Computation-based Science and Technology Research Center (CaSToRC), The Cyprus Institute, Nicosia, 2121, Cyprus}
\affiliation{Department of Mathematics, University College London, London, WC1H 0AY, UK}
\affiliation{Mathematics Institute, University of Warwick, Coventry, CV4 7EZ, UK}
\affiliation{Department of Mathematics, University of Hull, Hull, HU6 7RX, UK}
\affiliation{School of Mathematics, University of Birmingham, Birmingham, B15 2TT, UK}

\author{Panayiota Katsamba}\email{p.katsamba@cyi.ac.cy} \thanks{These authors contributed equally to this work}
\affiliation{Computation-based Science and Technology Research Center (CaSToRC), The Cyprus Institute, Nicosia, 2121, Cyprus}
\affiliation{School of Mathematics, University of Birmingham, Birmingham, B15 2TT, UK}

\author{Matthew D. Butler}\email{matthew.butler@ucl.ac.uk} \thanks{These authors contributed equally to this work}
\affiliation{Department of Mathematics, University College London, London, WC1H 0AY, UK}
\affiliation{Mathematics Institute, University of Warwick, Coventry, CV4 7EZ, UK}
\affiliation{School of Mathematics, University of Birmingham, Birmingham, B15 2TT, UK}

\author{Lyndon Koens}\email{l.m.koens@hull.ac.uk}
\affiliation{Department of Mathematics, University of Hull, Hull, HU6 7RX, UK}

\author{Thomas D. Montenegro-Johnson}\email{Tom.Montenegro-Johnson@warwick.ac.uk}
\affiliation{Mathematics Institute, University of Warwick, Coventry, CV4 7EZ, UK}
\affiliation{School of Mathematics, University of Birmingham, Birmingham, B15 2TT, UK}

\date{\today}

\begin{abstract}
We present an asymptotic theory for solving the dynamics of slender autophoretic loops and knots. Our formulation is valid for non-intersecting 3D centrelines, with arbitrary chemical patterning and varying (circular) cross-sectional radius, allowing a broad class of slender active loops and knots to be studied. The theory is amenable to closed-form solutions in simpler cases, allowing us to analytically derive the swimming speed of chemically patterned tori, and the pumping strength (stresslet) of a uniformly active slender torus. Using simple numerical solutions of our asymptotic equations, we then elucidate the behaviour of many exotic active particle geometries, such as a bumpy uniformly active torus that spins and a Janus trefoil knot, which rotates as it swims forwards. 
\end{abstract}


\maketitle

\section{Background}

Artificial microscale swimmers (microbots) are an exciting miniaturised technology, with promising applications in healthcare~\citep{nelson2010microrobots} and microfluidics~\citep{maggi2016self}. Examples of swimming microbots include helices with magnetic heads~\citep{Zhang2009b,GhoshFischer2009,Gaonanowire2010,Tottori2011barcrossshape}, magnetic sperm-like swimmers~\citep{Dreyfus2005}, oscillating-bubble powered capsules~\citep{bertin2015propulsion}, and phoretic microbots~\citep{paxton2004}. 
	
Phoretic microbots are active particles that propel via interaction with gradients of a field in the surrounding environment, such as electrical charge (electrophoresis)~\citep{nourhani2015self}, temperature (thermophoresis)~\citep{jiang2010,bickel2013}, or concentration of solute (diffusiophoresis)~\citep{golestanian2005}. 
In autophoretic propulsion, the microbot self-generates local gradients in the concentration of a surrounding solute fuel through differential reaction at its surface, often achieved by patterning the surface with a catalyst~\citep{paxton2004,GolestanianLiverpoolAjdari2007}.
	
Since the trajectory of autophoretic swimmers emerges from complex interactions between solute and fluid dynamics, it is dependent not only on the choice of catalytic patterning, but also on particle shape, and environmental factors such as position of domain boundaries. For instance, spherical Janus particles can follow the walls of microchannels~\citep{das2015boundaries}, straight rods with uniform coating can translate if their cross-section is non-uniform~\citep{schnitzer2015osmotic}, the end-profile of a slender rod can greatly affect its speed~\citep{Ibrahim_2017}, a particle comprising two linked, uniformly coated spheres of different sizes can reverse direction depending on their separation~\citep{michelin2015autophoretic}, uniformly coated spheres (which drive no fluid flow in isolation) can self-organise into mobile assemblies of multiple spheres~\citep{varma2018clustering}, Saturn rods (which pump flow in isolation) can self-organise into translating and rotating assemblies~\citep{wykes2016dynamic}, and a flexible uniform filament can pump fluid, translate, or rotate, depending on its centreline configuration~\citep{MontenegroJohnson2018Microtransformers,sharan2021fundamental}. It is noteworthy that the vast majority of these studies are characterised by rigid particles that are topologically equivalent to the sphere, with topologically nontrivial active particles receiving comparatively little study~\cite{schmieding2017autophoretic, baker2019shape}.

\begin{figure}
    \centering
    \includegraphics[width=\linewidth,viewport=40 590 570 750,clip]{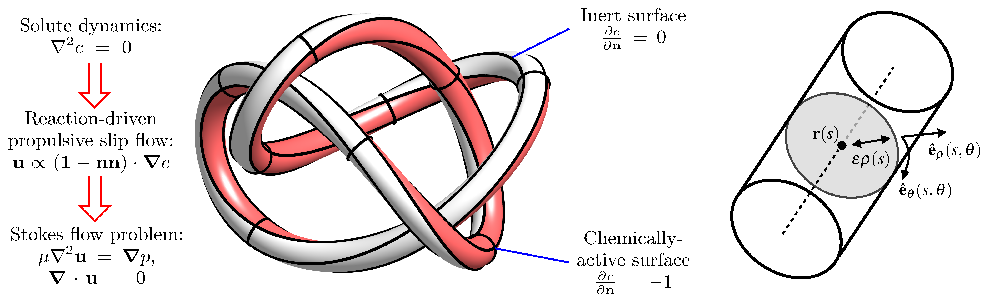}
    \caption{An example of a non-intersecting slender active loop (here, the $4_1$ ``figure of eight'' knot), to which our theory applies. Note the centreline is fully three-dimensional, and there is azimuthal variation in the chemical patterning. We may also consider loops with longitudinal variation in patterning, and varying cross-sectional radius.
    Right: a zoom in on a segment of the slender structure, highlighting a circular cross-section (shaded) at a given coordinate $s$, where the loop centreline is $\rc(s)$, the dimensionless cross-sectional radius is $\epsslend\crossradius(s)$, and the local coordinates in the plane of the cross-section are $\erho(s,\theta),\etheta(s,\theta)$.
    }
    \label{fig:LoopSchematic}
    
\end{figure}

Technological and theoretical advances are now allowing for the consideration of more complex systems. For example, an analytical solution for the dynamics of an axisymmetric autophoretic torus was derived by~\citet{schmieding2017autophoretic}.
These active tori were subsequently fabricated and shown to demonstrate translational swimming behaviour and cargo transport capabilities, as well as the ability to self assemble into dimers and trimers that could rotate~\cite{baker2019shape}. 
No analytical theory currently exists to study more general looped and knotted autophoretic microswimmers, and the developing capability to fabricate non-trivial phoretic designs paves the way for considering these more advanced geometries. It is not clear how these more complex shapes will interact and behave collectively; 
a key first step to understanding many-bodies is to study the individual particle dynamics.

Theoretical and computational frameworks exist that can solve for the phoretic swimming of particles of arbitrary shape. For example, \citet{lammert2016bypassing} showed that the phoretic swimming can be calculated directly from the surface solute flux by evaluating a surface integral, bypassing solving for the concentration field and slip velocity using the Lorentz reciprocal theorem. This approach has many advantages, since it is general and exact, and once the test solution is known in a given geometry, it can be reused for all chemical patternings on that geometry. However, in the case of slender bodies, calculating these integrals is often costly. 
Here,  we aim to use an approach tailored to slender objects that calculates the intermediate solutions for concentration and slip, since doing this admits natural extensions to multiple interacting swimmers \citep{varma2019modeling} and to fluid-structure interactions, once the relevant hydrodynamic surface tractions are calculated.
	
Looped and knotted shapes are often slender by design, that is, their cross-sectional radius is much smaller than their length. This slender limit allows asymptotic slender-body theories to be developed, which are computationally far less intensive than standard numerical techniques, such as boundary element methods \citep{Pozrikidis1992,Cortez05,montenegro2015regularised} which require a fine mesh to resolve the smallest scales across the entire body length. They also have the additional benefit of affording analytical solutions and insights. In the field of phoretic swimmers, a slender-body theory for straight electrophoretic rods was developed~\cite{yariv2008slender}, which was later extended to straight autophoretic rods with simple reaction kinetics~\cite{schnitzer2015osmotic}, and more complex reactions by Yariv~\cite{yariv2019self}. A matched asymptotics framework examined how end-shape and the cross-sectional profile of autophoretic rods affects swimming speed~\cite{Ibrahim_2017}, and detailed studies have shown the swimming behaviour of slender helices~\cite{Poehnl2021phoretic} and bent rods~\cite{ganguly2023going}. Recently came the development of Slender Phoretic Theory (SPT)~\cite{katsamba2020slender,katsamba2022chemically}, which is capable of analysing filaments with a general three-dimensional (curved, non planar) centreline and arbitrary cross-sectional radius and chemical patterning, and follows a previous asymptotic approach for slender filaments in viscous fluids~\cite{KoensLauga2018}.  

In this work, we present a Slender Phoretic Theory~\cite{katsamba2020slender} for non-intersecting, looped and knotted autophoretic filaments, such as that shown in Fig.~\ref{fig:LoopSchematic}, before applying our theory to derive new closed form analytical results for tori, as well as results for more complex designs. We begin with a broad discussion of the underlying theory. 

\section{Slender Phoretic Theory for active loops}

The dynamics of an autophoretic active particle are considered in the zero P\'{e}clet number limit, whereby the solute concentration at any instant is found from the solution to Laplace's equation, and the surface gradients of the resultant concentration field provides a slip flow boundary condition for the solution of a Stokes flow problem.

We derive the Slender Phoretic Theory for looped filaments (SPT Loops) by substituting the slender looped geometry into a boundary integral representation of Laplace's equation, and applying a matched asymptotic expansion in the small parameter, $\epsslend\ll1$, that quantifies the ratio of the maximum cross-sectional radius compared to the half-length of the filament centreline. The derivation of SPT Loops is involved, and given in full in Appendix~\ref{App_derivation}.

Broadly speaking, we shall find that the resulting solute concentration on the filament's surface is determined from an integral equation due to the given filament geometry and chemical patterning, with the slip flow found as its surface gradient. 
This slip flow can then be used as a boundary condition in a Slender Body Theory for Stokes flows to find the swimming and bulk fluid behaviour.

We will now present the main components of this theory, before proceeding to calculate results for various looped filaments with a range of chemical patterning. The presented results will be given in dimensionless form (as outlined below). 

\subsection{Governing equations}

We consider a phoretic swimmer in an infinite bath of fluid; the swimmer is chemically patterned so that it produces or depletes a solute due to a reaction occurring on its surface. Any gradients in the concentration of the solute by the swimmer's surface causes a slip flow in the fluid. If appropriately patterned, the swimmer can self-propel due to this slip flow.

We consider the limit of zero P\'eclet number~\citep{GolestanianLiverpoolAjdari2007}, where diffusion dominates advection of solute. 
This decouples the solute evolution from the fluid flow, and so the solute concentration, $c$, must obey Laplace's equation
\begin{equation}
	D\nabla^2c=0. \label{eq:Laplace}
\end{equation}

There are two major surface properties that characterise the behaviour of phoretic swimmers: the activity, $\activity$, which quantifies the generation or depletion of solute on the microbot's surface, and the mobility, $\mobility$, which encodes the generation of fluid slip velocities due to surface concentration gradients. These enter the governing equations as boundary conditions on the surface of the phoretic object 
\begin{equation}
	-D\no\cdot\boldsymbol{\nabla} c =\activity(\bv{x}) 
 ,  \label{flux_activity} 
\end{equation}
\begin{equation}
	\vslip = \mobility(\bv{x}) \left(\idmat - \no \no \right)\cdot\boldsymbol{\nabla} c
 ,\label{vslip_mobility}
\end{equation}
where $\no$ is the normal to the surface, pointing into the fluid. We choose $c$ to be the disturbance from a background concentration, so that $c\to0$ far from the surface, and, for simplicity, consider zeroth order reaction kinetics, where the activity $\activity$ is independent of concentration. 
This is the limit of zero Damk\"ohler number, where the diffusive transport rate dominates the reaction rate, valid for small particles, fast diffusion or slow reaction rates.
This assumption provides a simple starting point that could be extended in the future to more complex reaction behaviours \citep[e.g.][]{yariv2019self}. 

The fluid velocity, $\mathbf{u}$, is then governed by the Stokes flow equations
\begin{equation}
	\mu \nabla^2\bv{u} = \boldsymbol{\nabla}p, \quad \boldsymbol{\nabla}\cdot \bv{u} = 0,
\end{equation}
which is forced at the swimmer's boundary by the slip velocity
\begin{equation} 
    \bv{u} = \Uswim + \Omegaswim \times \surf + \vslip,
    \label{SurfaceVelocityBC}
\end{equation}
where $\Uswim, \Omegaswim$ are the swimmer's translational and rotational swimming velocities, respectively, and $\surf$ is the position vector of a point on the swimmer's surface.
These must be such that the swimmer is both force- and torque-free, i.e.~its surface traction $\bv{f}$ must satisfy
\begin{equation} \label{eq_ForceTorqueFree}
    \int_S \bv{f} ~\dd S = 0, \qquad
    \int_S \surf \times \bv{f} ~\dd S = 0,
\end{equation} 
where the integral is taken over the surface of the phoretic swimmer.

For a filament with total arclength $2\length$ and maximum cross-sectional radius $\maxcrossradius$, we non-dimensionalise length by $\length$, activity by a typical activity $[\activity]$, and concentration by a typical concentration 
$[c]\equiv[\activity]\maxcrossradius/D$. 
For a typical mobility scale, $[\mobility]$, 
we then find a typical phoretic velocity scale 
$\vslip \sim [\mobility][\activity] \maxcrossradius/ (D\length)$. 
We use these scales to non-dimensionalise our slender phoretic loops: all quantities are henceforth dimensionless, unless otherwise stated.

\subsection{Looped filament geometry}

We capture the geometry of the slender loops in a similar manner to previous work on slender filaments~\citep{KoensLauga2018, KoensLauga2017_AnalytSolSRT, katsamba2020slender}. The slender loop has a centreline, $\rc(s)$, which is parametrised by its arclength $s\in\left[-1,1\right]$. We denote the vector from the centreline at arclength $\sdum$ to the centreline at $s$ by $\bvRo(s,\sdum) = \rc(s)-\rc(\sdum)$. The centreline has a tangent $\tanhat(s)$, normal $\norhat(s)$, and binormal $\binorhat(s)$ that satisfy the Serret-Frenet equations, 
\begin{equation}\label{SerretFrenetEqns}
 \dds{\tanhat}= \curvature \norhat, \quad 
 \dds{\norhat}= -\curvature\tanhat + \torsion \binorhat, \quad 
 \dds{\binorhat}=-\torsion \norhat,  
\end{equation}
where $\curvature,\torsion$ are the curvature and torsion of the centreline, respectively.
Note that, unlike a filament with free ends, a looped filament has periodicity in $s$, 
so that the two ends are joined at $s=\pm 1$, i.e.~$\rc(-1) = \rc(1)$.

We focus on filaments with a circular cross-section. Cross-sections are perpendicular to the centreline tangent, as shown in the schematic in Fig.~\ref{fig:LoopSchematic} The cross-sectional radius is captured by $\epsslend \crossradius(s)$, where $\epsslend=\maxcrossradius/\length$ is the slenderness of the filament and $0<\crossradius(s)\leq1$. Note the assumption that  $\crossradius(s)$ remains non-zero everywhere along the centreline, so that the filament is a continuous closed loop. 

The surface of the filament is then parametrised by its arclength, $s$, and the azimuthal angle of the cross-section, $\theta\in\left[0,2\pi\right]$, through the relation 
$\surf(s,\theta) = \rc(s) + \epsslend\crossradius(s) \erho  (s,\theta). \label{surf_param}$
Here, $\erho(s,\theta)$ is the local radial unit vector perpendicular to the centreline tangent, and $\etheta(s,\theta)$ is the corresponding azimuthal unit vector (around the slender direction), as illustrated in Fig.~\ref{fig:LoopSchematic}. Note that the curve on the surface with $\theta=0$ can be chosen arbitrarily, and so we also define a fixed coordinate frame angle $\dtheta(s,\theta)=\theta-\thetators(s)$, that tracks the azimuthal angle compared to $\thetators(s)$, which rotates with the torsion of the centreline, $\partial\thetators/\partial s = \torsion$. With this definition, $\dtheta=0$ occurs where the surface normal aligns with the Serret-Frenet normal $\norhat(s)$, so that $\erho(s,\theta) = \cos\dtheta(s,\theta)\norhat(s) + \sin\dtheta(s,\theta)\binorhat(s)$ and $\etheta(s,\theta) = -\sin\dtheta(s,\theta)\norhat(s) + \cos\dtheta(s,\theta)\binorhat(s)$. This angle parametrisation is convenient as it removes the influence of the line torsion, $\tau$, from the leading order equations \cite{KoensLauga2018}.

\subsection{Slender expansion}

For slender filaments, with $\epsslend\ll1$, we look for solutions in terms of an asymptotic expansion in the slenderness. For example, the concentration field shall be expanded as
\begin{equation}
    c(s,\theta) = \zerothorder{c} + \epsslend \firstorder{c} + O(\epsslend^2).
\end{equation}
Similarly, quantities to be calculated --- such as the fluid velocities and swimming speed --- shall be expanded in $\epsslend$, with a bracketed superscript denoting the order of the relevant term. 

The derivation of SPT Loops follows the same procedure as for open-ended filaments~\cite{katsamba2020slender}. However, due to the key change in topology between looped and unlooped filaments, a complete re-derivation of the theory is required for full academic rigour, rather than simply applying or extending the existing theory. 

This theory is derived systematically from a boundary integral representation of Laplace's equation for the solute diffusion. Performing a matched asymptotic expansion allows us to transform the surface integral equation into a line integral formula for evaluating the concentration on the filament surface, thereby significantly simplifying its calculation.

In particular, the surface solute concentration is determined from two main contributions: a non-local integral contribution due to the geometry and the cross-sectionally-averaged activity, which are integrated over the entire centreline of the filament (the outer region), and a local contribution due to the nearby region where the thickness of the filament is important (the inner region).

Importantly, the slip flow that drives the bulk fluid motion is generated by gradients in the surface solute concentration, and concentration gradients in the azimuthal direction (around the small cross-section of the filament) are enhanced relative to gradients along its length. It is therefore often necessary to determine the next order contribution to the concentration as this can impact the swimming behaviour at leading order.

\subsection{Slender Phoretic Theory for Loops}

We summarise the main equations of SPT Loops in Eqs.~\eqref{eq:Conc_GenActivity}--\eqref{eq:SlipVelocity} below, with the derivation provided for completeness in Appendix~\ref{App_derivation}. In particular, we give an expression for the leading order surface solute concentration for an arbitrary activity profile, $\activity=\activity(s,\theta)$, and the first two orders of the concentration when the activity is axisymmetric, $\activity=\activity(s)$, as well as the resulting slip velocity. 

\paragraph{General activity, $\activity=\activity(s,\theta)$}
\begin{multline}
        2\pi \zerothorder{c} (s,\theta) 
        = \frac{1}{2}	~ \int\limits_{-1}^{1}\! \left[
        \frac{\crossradius(s+q)\langle\activity(s+q)\rangle}{ |\bvRo(s,s+q)|}
        -\frac{\crossradius(s)\langle\activity(s)\rangle }{|q|}  \right]
        \dd q   
        +
        \frac{1}{2} \crossradius(s)	\langle\activity(s) \rangle \log 
        \left( \frac{1}{\epsslend^2  \crossradius^2(s)}\right) \\  
        -
        \crossradius(s)	\int_{-\pi}^{\pi}\activity(s,\thetadum) 
        \log\left[1-\cos(\theta-\thetadum)\right] \dd \thetadum + O(\epsslend).
        \label{eq:Conc_GenActivity}
\end{multline}

\paragraph{Axisymmetric activity, $\activity=\activity(s)$}
\begin{align}
        \zerothorder{c}(s)
        =& 
        \frac{1}{2} \int_{-1}^{1} \left[
        \frac{\crossradius(s+q)\activity(s+q)}{ |\bvRo(s,s+q)|} 
        -\frac{\crossradius(s)\activity(s)}{|q|} \right]
         \dd q 
        +	\frac{1}{2}
        \crossradius(s)\activity(s)\log\left(\frac{4 }{\epsslend^2 \crossradius^2(s)}\right) 
        \label{eq:ZerothOrderConc_AxiActivity}
        \\
        \firstorder{c}(s,\theta)
        =& 
             \frac{1}{2} \crossradius^2(s) \curvature(s) \activity(s)	
            \cos\dtheta(s,\theta)  \left[\log\left(\frac{4}{\epsslend^2 \crossradius^2(s)}  \right)
            - 3\right]
        \nonumber
        \\
        &\qquad - \crossradius(s) \int\limits_{-1}^{1}\!
        \left[ 
        \frac{\crossradius(s+q)\activity(s+q)}{ |\bvRo(s,s+q)|^3}    
              \bvRo(s,s+q)    \cdot\erho(s,\theta)
          +  \frac{\crossradius (s)\curvature(s)\activity(s)\cos\dtheta(s,\theta)}{2|q|}\right]\dd q.   
          \label{eq:FirstOrderConc_AxiActivity}
\end{align}

\paragraph{Slip velocity}
\begin{align}
        \frac{1}{\mobility}	\vslip(s,\theta) 
        = \underbrace{ \etheta \frac{1}{\epsslend \crossradius(s)}\partial_\theta \zerothorder{c}}_{O(1/\epsslend)}
        +\underbrace{\left[ \etheta \frac{1}{  \crossradius}\partial_\theta \firstorder{c} 
        + \tanhat \partial_s \zerothorder{c}\right]}_{O(1)} + O(\epsslend). 
        \label{eq:SlipVelocity}
\end{align}

Note that the general activity case, Eq.~\eqref{eq:Conc_GenActivity}, has been integrated over $\theta$, which gives rise to the factor $2\pi$ on the left-hand side, as well as the $\theta$-averaged terms, $\langle \cdot \rangle = \int_{-\pi}^{\pi} \cdot ~\dd \theta$, on the right-hand side. Only the leading order concentration is shown for non-axisymmetric activities since, in these cases, we would typically expect a $\theta$-dependence in the leading order surface concentration, $\zerothorder{c}$, which by itself gives the leading order behaviour of the slip velocity at $O(1/\epsslend)$, as seen from Eq.~\eqref{eq:SlipVelocity}.  

The effect of changing from open-ended to looped filaments in slender body theories has been previously achieved in the context of viscous hydrodynamics of ribbons, called Slender Ribbon Theory (SRT)~\cite{KoensLauga2017_AnalytSolSRT}. 
Correspondingly, we find the expansion of Slender Phoretic Theory to looped filaments takes a similar form to the free-end SPT case, but with $\sdum$ replaced by $s + q$ in the integrals, where $q$ is now the integration variable, whilst also removing the $(1-s^2)$ factor from inside the logarithmic term. This substitution $\sdum\to q$ describes a looped system as the integration becomes independent of the choice of origin ($ s= 0$). We emphasise that, although similar effects are observed in other similar theories such as SRT, a full re-derivation (see Appendix~\ref{App_derivation}) was required to confirm this.

\subsection{Fourier series representation} \label{sec:FourierSeries}

At any given arclength, $s$, all properties on the surface of the filament must be periodic in $\theta$; this periodicity suggests using a Fourier decomposition in the azimuthal angle~\cite{katsamba2022chemically}. We can therefore expand the (known \textit{a priori}) activity in terms of Fourier modes as 
\begin{equation}
    2\pi \crossradius(s) \activity(s,\theta) 
    = \Ao(s) 
    + \sum_{n=1}^{\infty} \left[\Acoeffcosmode{n}(s)\cos n\thetamode 
    + \Acoeffsinmode{n}(s)\sin n\thetamode\right], \label{Activity_Fourier}
\end{equation}
where $\Ao = \crossradius(s) \langle\activity(s)\rangle $ for $\langle\activity(s)\rangle = \int_{-\pi}^{\pi} \activity(s,\theta) \dd \theta$.
Similarly, we expand the (to be determined) solute concentration and slip velocity as
\begin{align}
    2\pi c(s,\theta) &= c_0(s) 
    + \sum_{n=1}^{\infty} 
    \left[ \ccoeffcosmode{n}(s)\cos n\thetamode 
    +  \ccoeffsinmode{n}(s)\sin n\thetamode \right],   \label{Conc_Fourier} 
    \\
    2\pi \vslip(s,\theta) &= \vslipzeromode(s) 
    + \sum_{n=1}^{\infty} 
    \left[ \vslipcoeffcosmode{n}(s)\cos n\thetamode 
    +  \vslipcoeffsinmode{n}(s)\sin n\thetamode \right]. 
    \label{Vslip_Fourier}
\end{align}
From the known Fourier coefficients for the activity, the aim is then to calculate the Fourier coefficients for the concentration field and slip velocity.

Substituting these Fourier expansions into the leading order expression for the surface solute concentration for a general activity, 
Eq.~\eqref{eq:Conc_GenActivity},
and noting that
\begin{align}
  &\int_{-\pi}^{\pi}\log\left[1-\cos\phi\right]\dd \phi =  -2\pi \log 2, \\
  & \int_{-\pi}^{\pi}\cos(n\phi)\log\left[1-\cos\phi\right]\dd \phi =  \frac{-2\pi}{n}, \qquad\, \text{for }n>0, \\
  &  \int_{-\pi}^{\pi}\sin(n\phi)\log\left[1-\cos\phi\right]\dd \phi =  0, \qquad\qquad \text{for }n\geq0,
\end{align}
we find that the Fourier coefficients for the solute concentration on the filament surface are given by
\begin{align}
        2\czeroorderzeromode(s) 
           = & 
        	\int_{-1}^{1} \left[
        \frac{\Ao(s+q)}{ |\bvRo(s,s+q)|} 
        -\frac{\Ao(s)}{|q|} \right]
         \dd q  
        +
        \Ao(s)\log\left(\frac{4 }{\epsslend^2 \crossradius^2(s)}\right), \label{c_0_mode_fourier}
        \\
        \czeroordercoeffcosmode{n}(s)  =&  \frac{1}{n}\Acoeffcosmode{n}(s), \label{eq:cos_mode_fourier}
        \\
        \czeroordercoeffsinmode{n}(s) =&   \frac{1}{n}\Acoeffsinmode{n}(s). \label{eq:sin_mode_fourier}
\end{align}
This decomposition into Fourier modes is often useful in practice, and will be applied in the applications to follow.

\subsection{Swimming}

Having resolved the effect of activity on the surface concentration field and the concomitant slip velocity from the SPT Loops equations, it remains to determine the motion of the swimmer. In the lab frame, the fluid velocity at the swimmer's surface is composed of the aforementioned slip velocity and the rigid body motion of the swimmer
\begin{equation}
    \bv{u} (s,\theta) = \Uswim + \Omegaswim \times \rc(s) + \vslip(s,\theta)
\end{equation}
where $\Uswim$ is the leading-order swimming velocity and $\Omegaswim$ is the leading-order angular velocity, both of which are to be determined. These velocities must be such that the swimmer is force- and torque-free, so that the surface traction, $\bv{f}(s,\theta)$, satisfies Eq.~\eqref{eq_ForceTorqueFree}.

From the slip velocity of the SPT Loops equations, we can then calculate the swimming motion of the phoretic loops using any appropriate Slender Body Theory for Stokes flows (SBT). We choose to use the SBT of~\citet{KoensLauga2018}, which relates the Fourier modes of the surface fluid velocity to the surface tractions via an integral equation, combined with force- and torque-free conditions. This method is outlined in~\citet{katsamba2020slender}.

\section{Numerical implementation of Slender phoretic theory}

Although in some special cases, the SPT equations can be solved analytically~\citep{katsamba2022chemically}, in general, they must be solved numerically for arbitrary geometry and chemical patterning. Alongside this work, we provide a MATLAB code that can numerically calculate the solute concentration, slip velocity and swimming behaviour of a filament for a given filament centreline, radius and activity~\cite{SPTcode}. This code has been written to solve for the looped filaments presented in this work, as well as the previously-considered case of open-ended filaments~\citep{katsamba2020slender,katsamba2022chemically}. This unoptimised code typically runs within seconds in MATLAB on a laptop. 

The inputs are the geometry via the centreline, $\rc$, cross-sectional radius, $\crossradius$, and slenderness, $\epsslend$, as well as the activity over the surface, $\activity$. Our code first calculates the arclength coordinates and renormalises to a total arclength 2 (if required), before calculating the tangent, normal and binormal. From these, a quadrature mesh is generated in the arclength direction, $s$. Variations in the activity in the cross-sectional direction, $\theta$, are decomposed into a (truncated) Fourier series, as described in \S\ref{sec:FourierSeries}, with cosine and sine coefficients stored as real and imaginary parts of the vector of modes, respectively. 

The line integral for the ($\theta$-averaged) leading-order surface concentration, $\zerothorder{c}$, is then evaluated via quadrature, with higher order Fourier modes simply determined as multiples of the corresponding activity modes. If the activity is axisymmetric, then the next-order contribution, $\firstorder{c}$, is also calculated via quadrature. 

The resulting leading-order slip velocity is calculated by taking derivatives of the concentration. Derivatives in the longitudinal direction, $s$, are calculated using a spline interpolant, while $\theta$-derivatives are calculated by appropriately updating the Fourier series.

The swimming behaviour is determined from the slender body theory of~\citet{KoensLauga2018}, using the same method detailed in~\citet{katsamba2020slender}. This involves discretising the slender body integral equations for the known slip velocities, given as line integrals of the unknown surface tractions, as well as the force- and torque-free integral conditions, to reduce these to a matrix equation $A\bv{x}=\bv{b}$ where $\bv{x}$ is a vector of surface tractions and the swimming velocities, and $\bv{b}$ is a vector of the $\theta$-averaged slip velocities. The matrix $A$ is determined from the geometry of the slender object. This linear equation is inverted to determine $\bv{x}=A^{-1}\bv{b}$, and hence the swimming behaviour of the slender filament.

We now use our theory to analyse the dynamics of key examples of slender chemically active loops and knots. As it is a commonly-used chemical patterning, both in theory and practice, we will mostly consider examples with ``Janus" designs, whose surface is comprised of uniformly active and inert regions. For simplicity, we shall also apply a uniform mobility over the whole surface. However, note that the theory is valid far beyond this, and can handle general variations in surface chemical patterning.  All results are presented for active regions that deplete the solute (negative activity), with slip flows directed up the chemical gradient (positive mobility); as such, we may typically expect the net swimming behaviour to be towards active regions.

\section{The torus}

\begin{figure}
\subfloat[Glazed torus]{\includegraphics[width=0.3\textwidth,viewport=300 300 850 550,clip]{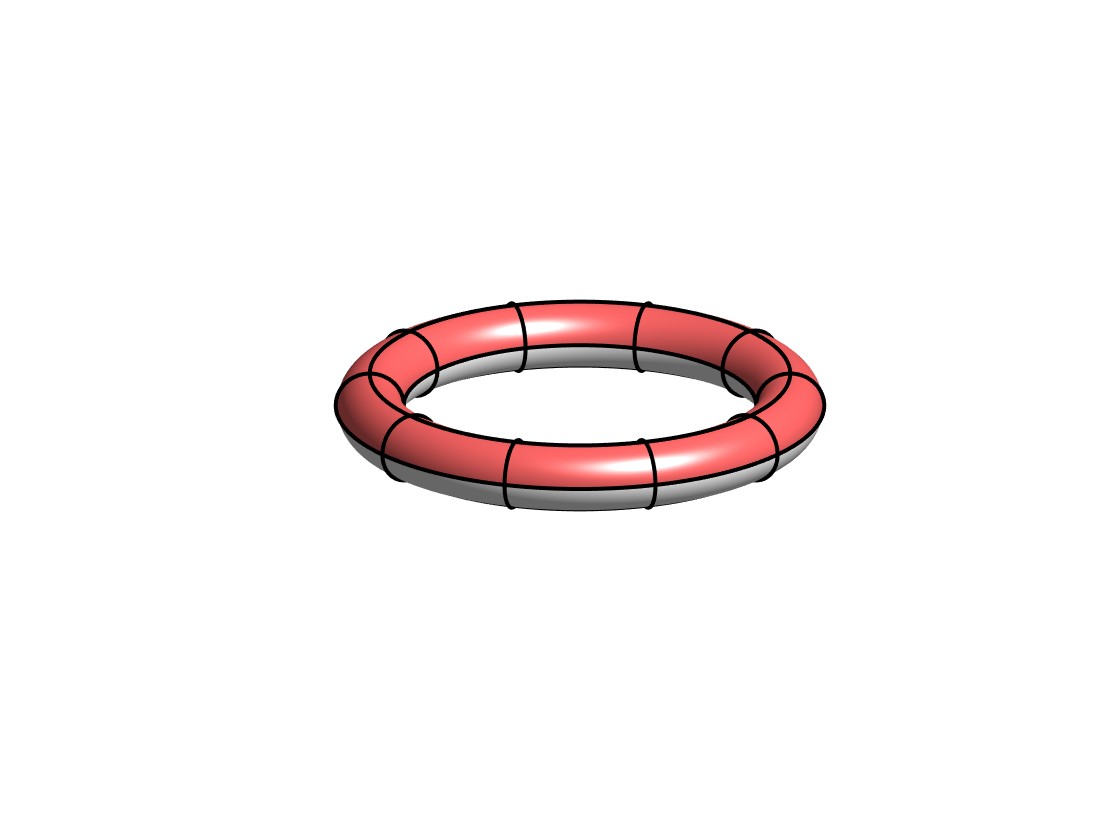}}
\hspace{0.5cm}
\subfloat[Dunked torus]{\includegraphics[width=0.3\textwidth,viewport=300 300 850 550,clip]{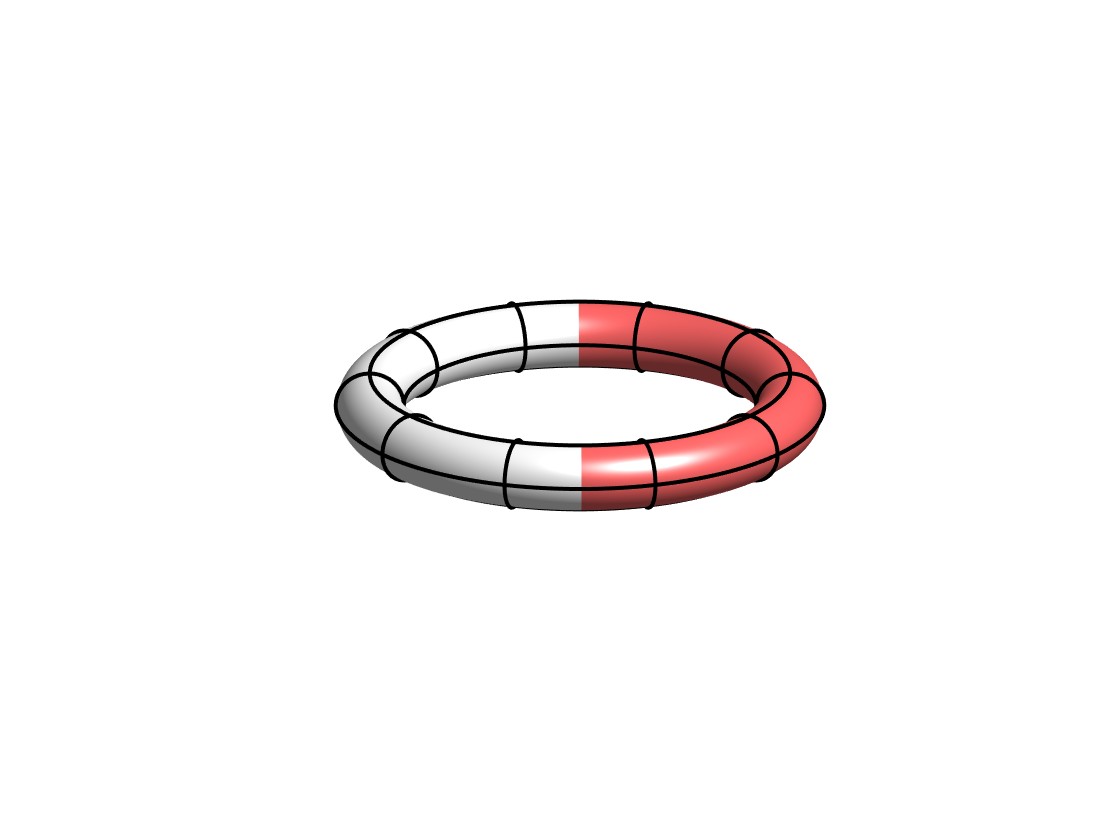}}
\hspace{0.5cm}
\subfloat[Uniform torus]{\includegraphics[width=0.3\textwidth,viewport=300 300 850 550,clip]{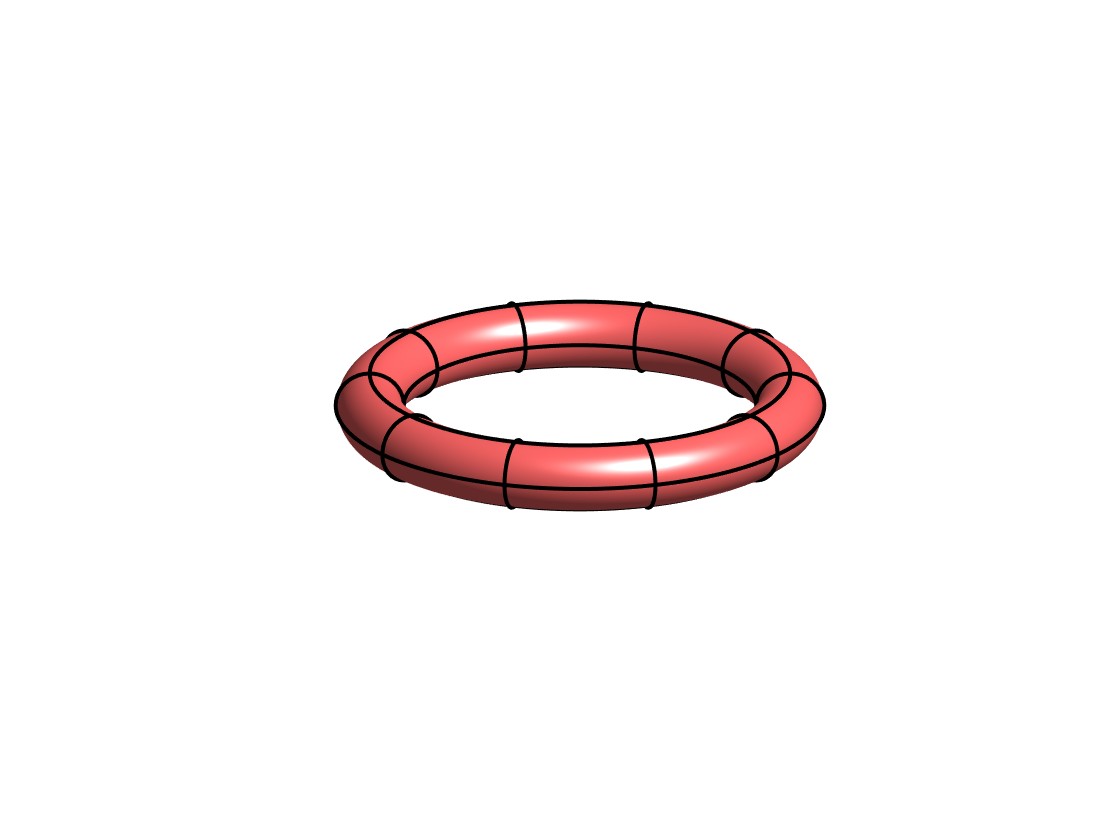}}
\caption{The active slender tori considered herein. Red regions represent active surfaces, while white regions represent inert surfaces. The aspect ratio of the swimmers has been increased for easier visualisation.
(a) ``Glazed'' activity profile with azimuthal variation swims along its symmetry axis. 
(b) ``Dunked'' activity profile with local axisymmetry swims in-plane.
(c) The uniformly active torus does not swim, but acts as a pump (stresslet). }
\label{fig:Donuts}
\end{figure}

The simplest looped filament is the torus (an unknot). The earliest conception of a swimming microscale torus was given by~\citet{purcell77}, where the proposed method of locomotion was ``tank treading''. Slender body theories have previously been applied to tori to find the velocity-drag relations for translation and rotation along different axes~\cite{johnson1979torus}, which were subsequently extended to include the effect of electrophoresis~\cite{solomentsev1994electrophoresis}.~\citet{leshansky2008surface} later found a series solution for the swimming of a ``squirming'' torus, valid for any axisymmetric distribution of surface slip velocity. This approach was used by~\citet{schmieding2017autophoretic} to derive a solution for the phoretic propulsion of any axisymmetrically chemically-patterned torus, which led to the first experimental realisation of toroidal active particles by~\citet{baker2019shape}.

Here, we demonstrate some of the power and flexibility of our slender theory by deriving analytical results for archetypal chemically-active tori, illustrated in Fig.~\ref{fig:Donuts}: two types of half-coated tori that swim and have been previously manufactured~\cite{baker2019shape} (the ``glazed" torus and ``dunked" torus), and a uniformly active torus (the ``uniform" torus). The uniformly active torus does not swim (by symmetry), but rather acts as a pump via confinement effects; remarkably, we are able to derive the strength of this pumping (the ``stresslet'') analytically in the slender case.

\subsection{Geometry}

We consider a torus with a circular centreline that lies in the $(x,y)$-plane, with a constant cross-sectional radius $\crossradius(s)\equiv1$. 
The torus centreline can be parametrised by the arclength, $s$, such that
\begin{equation}
    \bv{r}(s) =  \begin{pmatrix} x(s) \\ y(s)  \end{pmatrix} = 
    \frac{1}{\pi}\begin{pmatrix} \sin (s \pi) \\
    1-\cos(s \pi)
    \end{pmatrix}. 
    \label{toruscentrelinegeom}
\end{equation}
Then the distance between two points on the centreline is
\begin{equation}
    |\bvRo(s,s+q)|   
    =   \frac{2}{\pi} \left| \sin \left( \frac{\pi q}{2} \right) \right|.
\end{equation}
Note also that the planar centreline has no torsion, hence 
$\thetators=0$, and $\thetamode = \theta$, and that the curvature is constant, taking a value $\curvature=\pi$. Here, we define $\theta=0$ as the innermost edge of the torus, with $0<\theta<\pi$ on the `top' surface. The normal $\norhat$ lies within the plane of the centreline, pointing towards the centre of torus, and $\binorhat$ is directed upwards out of the plane of the centreline, $\binorhat=\bv{e}_z$. 
 
\subsection{Glazed and dunked tori}

Swimming at the small scale requires symmetry-breaking~\cite{purcell77}.
A simple mechanism to induce swimming on a highly symmetric particle, such as a torus, is to vary the activity over its surface. The resulting asymmetric concentration gradients can then generate a net slip flow past the torus that propels it forwards. 

We consider two main types of activity variation, depicted in Fig.~\ref{fig:Donuts}: a ``glazed" torus with activity that varies in the slender (azimuthal) direction (Fig.~\ref{fig:Donuts}a), and a ``dunked" torus which varies around the loop but is locally axisymmetric (Fig.~\ref{fig:Donuts}b). By symmetry, the glazed torus must swim out-of-plane and the dunked torus swim in-plane, both with no rotation.

We calculate the swimming velocity of the glazed torus analytically by decomposing into Fourier modes in the azimuthal direction. Here, we present results for a sharp transition in the activity (Janus) with one half inert and the other half active, i.e.~$\activity=-1$ in one half and $\activity=0$ in the other, with uniform mobility $\mobility$. In Appendix~\ref{sec:GlazedSinusoid}, we also present the calculation of the swimming for a smoothly-varying sinusoidal activity around the azimuthal direction. For the dunked torus, we determine analytic expressions for the solute concentration and the slip velocity, but resort to calculating the resulting swimming numerically.

\subsubsection{Glazed torus}
For a glazed torus with a Janus profile, that is one half is active and the other inert, the activity is given by 
\begin{align}
      \activity(s,\theta) = 
      \begin{cases} 
     -1, \quad 0    <\theta  < \pi, \\
     0, \quad -\pi < \theta < 0,  
      \end{cases}
  \end{align}
and can be decomposed into Fourier modes as
\begin{align} 
    2\pi  \activity(s,\theta) 
    = -\pi 
    +
     2 \sum_{n=1}^{\infty}   \frac{1}{n}  \left[(-1)^n-1\right]\sin n\theta   .
\end{align}
  
The leading-order surface concentration is then 
 \begin{align} \label{eq:GlazedTorusConc}
       2\pi \zerothorder{c}(s,\theta) 
    =&
    	-\pi \log\left(\frac{8 }{\epsslend \pi   }\right)   
    	+ 2  \sum_{n=1}^{\infty} \frac{1}{n^2} \left[(-1)^n-1\right]\sin n\theta ,
 \end{align} 
and the slip velocity can be calculated to be
 \begin{align}
\vslip(s,\theta)    
=& \frac{\mobility}{ \pi \epsslend  } \etheta(s,\theta) 
\sum_{n=1}^{\infty}  \frac{1}{n} \left[(-1)^n-1\right] \cos n\theta   +O(1) 
\nonumber \\
 =& -\frac{\mobility}{ \pi \epsslend  } \mbox{arctanh}[\cos(\theta)] \etheta(s,\theta) +O(1).
\end{align}
Note that this is the same result as the azimuthally-Janus slender phoretic filaments that was found in~\cite{katsamba2022chemically}.

This can be integrated around a cross-section to find the zeroth Fourier mode for the slip velocity
\begin{equation}
    \vslipzeromode(s) = -\frac{2\mobility}{\epsslend} \bv{e}_z,
    \label{vslip_zeromode_glazedJanus} 
\end{equation}
where we have used the fact that, since the torus is planar, the binormal points along the symmetry axis, $\binorhat=\bv{e}_z$.

To calculate the swimming speed, $\Uswim$, we first need to consider the surface tractions, $\bv{f}$, acting on the torus. We focus on a torus that has a uniform mobility, $\mobility=$ constant. Using symmetries of Stokes flows on a torus with the given slip, we find that the leading-order traction must act in the $z$-direction only, $\bvfo(s) = f_0 \bv{e}_z$, and $f_0$ must be constant.  The slender body theory of~\citet{KoensLauga2018} then gives the Fourier coefficients for the surface fluid velocity in terms of the surface tractions. In our case, we find that the zeroth Fourier mode for the surface fluid velocity, $\bvUo$, which is the only mode important to determine the translational motion, is
\begin{align}
4 \bvUo(s)
&=f_0 \bv{e}_z  \left[ 1+  \log
\left( \frac{64}{ \epsslend^2 \pi^2  }  \right)        \right] + O(\epsslend).
\end{align}

Since $f_0$ is constant, the condition of total zero force on the swimmer is only satisfied when $f_0=0$, and so the surface fluid velocity is zero, $\bvUo = \bv{0}$. But this fluid velocity is composed of the slip velocity and the swimming velocity, $\bvUo = \bvUo^{swim} + \vslipzeromode$, and so the swimming velocity can be simply determined as
\begin{equation}
    \Uswim =  \frac{ \mobility}{   \pi \epsslend  } \bv{e}_z.
    \label{eq:JanusGlazedSpeed}
\end{equation}
where we note the change in factor of $2\pi$ from Eq.~\eqref{vslip_zeromode_glazedJanus} due to the initial azimuthal averaging to obtain the Fourier modes.
This swimming speed for the glazed torus is plotted in Fig.~\ref{fig:MainResults}a as function of $\epsslend$, for comparison with later results. 

Note that, although the swimming speed appears to diverge as $\epsslend\to0$, the velocity has been scaled by a factor $AM\epsilon/D$ when rendered dimensionless (for asymptotic convenience) and so the fully dimensional speed is in fact constant with respect to slenderness. 

\subsubsection{Dunked torus}

We now consider a second example of a Janus torus, the dunked torus, which is patterned in a perpendicular manner to the glazed torus, so that the activity is uniform over any cross-section, as illustrated in Fig.~\ref{fig:Donuts}b. 
This torus again has a constant cross-section, $\crossradius(s)=1$, but now the axisymmetric activity profile is given by
 \begin{align}
     \activity(s) = 
     \begin{cases} 
     -1, \quad 0 < s < 1, \\
     0, \quad -1 < s < 0. 
     \end{cases}
 \end{align}
 
We split the results into active and inert regions, and calculate the integrals separately in each region. These integrals can be evaluated analytically, using the fact that $\dd/\dd x ~[\log\tan(\pi x/4)] = \pi/[2 \sin(\pi x/2)]$. 
For notational compactness, we define the functions 
\begin{equation}
    \tanfunction(x) = \tan \left(\frac{\pi|x|}{4}\right),
    \qquad \qquad
    \sinfunction(x)=\frac{1}{\sin \left(\frac{\pi|x|}{2}\right)}.
\end{equation}
In the active region, $s>0$, the concentration profile is given by
\begin{equation}
    c(s,\theta) 
    = -\frac{1}{2}  \log \left[  \frac{ 64  }{ \pi^2 \epsslend^2 } \tanfunction(1-s) \tanfunction(s)
    \right] 
    - \epsslend \frac{\pi}{2} \cos\theta \left\{ 
    \log \left[  \frac{ 64 }{ \pi^2 \epsslend^2 } \tanfunction(1-s) \tanfunction(s)
    \right]
    - 3 
     \right\}
    + O(\epsslend^2),
\end{equation}
whereas in the inert region, $s<0$, it is
\begin{align}
    c(s,\theta) = -\frac{(1+\pi\epsslend\cos\theta)}{2} \log \left[ 
    \frac{\tanfunction(1-s)}{\tanfunction(s)} 
    \right] + O(\epsslend^2).
\end{align}

The resulting slip velocity in the active region, $0<s<1$, is therefore given by
\begin{align}
    \vslip = -\frac{\pi\mobility}{2} \Bigg[
    \left( 
    \frac{\sinfunction(s)-\sinfunction(1-s)}{2}
    \right)\tanhat 
    - \left( \log\left[ \frac{64}{\pi^2\epsslend^2}
    \tanfunction(1-s)\tanfunction(s)
    \right] -3 \right) \sin\theta \etheta
    \Bigg],
\end{align}
and in the inert region, $-1<s<0$, it is given by
\begin{equation}
    \vslip 
    = -\frac{\pi\mobility}{2} 
    \left[  \left( 
    \frac{\sinfunction(s)-\sinfunction(1-s)}{2}
    \right)\tanhat 
    -  \log\left[ 
    \frac{\tanfunction(1-s)}{\tanfunction(s)}
    \right]  \sin\theta \etheta
    \right].
\end{equation}
Note that these results are functionally the same as found for a slender circular arc with an active middle section~\cite{katsamba2022chemically}.

We numerically calculate the swimming behaviour (from the geometry and activity, with the numerically calculated concentration and slip agreeing closely with the presented analytical results).
Results for the swimming speed are shown in Fig.~\ref{fig:MainResults}b (presented later in the text for comparison with other results), showing a near constant behaviour for the dimensionless in-plane swimming speed as a function of the slenderness, $U_{\mathrm{sw}}\approx1.4$, with no rotation. 
We also consider the effect of varying the activity coverage from completely inert to completely active. As shown in Fig.~\ref{fig:MainResults}c, the maximum swimming speed is obtained for the half-active dunked torus, with no swimming when completely active or completely inert.

\subsection{Uniform tori}
\label{sec:UniformTorus}

The simplest active torus has uniform activity, $\activity(s) = -1$ (as shown in Fig.~\ref{fig:Donuts}c). The inside walls of this ``uniform'' torus have enhanced solute depletion compared to the outside, due to geometric confinement, creating a surface concentration gradient that drives a slip flow from inside to outside. 

The symmetry of the uniform torus means that it will not swim, but, in the far-field, the fluid velocity is given by a straining flow which has the general form
\begin{align}
    \bv{u}(\bv{y}) = 
    \frac{3(\bv{y}\cdot\stresslet\cdot\bv{y})\bv{y}}{8\pi\mu|\bv{y}|^5},
\end{align}
where $\stresslet$ is a second-rank tensor called a stresslet. Our aim is to find an expression for this stresslet from our Slender Phoretic Theory for Loops.

We begin by noting that the integral in the $\zerothorder{c}$ term can be evaluated as
\begin{equation}
 	 \int_{-1}^{1} \left[
\frac{\pi/2}{| \sin \left(\pi q/2 \right)|}
-\frac{1 }{|q|} \right]
 \dd q
 =
    \log
\left[  \frac{ 16 }{ \pi^2    }  \right] ,  \label{circintegralresult}
\end{equation}
and that the integrand in the expression for $\firstorder{c}$ can be simplified using
\begin{align}
    \bvRo(s,s+q) \cdot \erho(s,\theta) 
    &=  \frac{\cos (\pi q) -1}{\pi} \cos\theta  
    = - \frac{2}{\pi}    \sin^2 \left( \frac{\pi q}{2}\right) \cos\theta.
\end{align}
From these, we find that the concentration field on the surface of the torus is given by
\begin{align} \label{eq:UniformDonut_Conc}
	c(s,\theta)
  = 
  -\logfactor
- \epsslend \pi\cos\theta \left[ 
\logfactor
- \frac{3}{2} \right]  
   + O(\epsslend^2),   
\end{align}
where $\logfactor = \log(8/\epsslend\pi)$. This differs from the surface concentration of a uniformly active circular arc in the limit of closing the loop~\cite{katsamba2022chemically}, since the end effects are important there; however, this result is equivalent to setting $s=0$ everywhere in the SPT result for a full circular arc (since there is no obvious choice of origin once the loop is closed).

Since the uniform torus has an axisymmetric activity, the leading-order slip velocity will be $O(1)$, with contributions from both the leading-order surface concentration gradient in the $s$-direction and next-order surface concentration gradient in the $\theta$-direction. However, the leading-order surface concentration is uniform. Therefore, the leading term in the slip velocity is
\begin{align}
\frac{1}{\mobility}	\vslip(s,\theta)   
&=    \etheta \frac{1}{  \crossradius(s)}\partial_\theta \firstorder{c} 
+ O(\epsslend)  
=\etheta 
 \pi  \sin\theta  
\left[
\logfactor
    - \frac{3}{2} 
\right]
   + O(\epsslend). 
\end{align}
This can be expanded in terms of the tangent, normal and binormal vectors that represent our slender curve
and
integrated in $\theta$ to find the zeroth Fourier mode for the slip velocity
\begin{align}
    \vslipzeromode(s) 
    &= \int_{-\pi}^{\pi}	\vslip(s,\theta)   \dd \theta 
   = -\norhat(s) \pi^2 \mobility 
\left[
\logfactor
    - \frac{3}{2} 
\right].   \label{Uo_uniformdonut}
\end{align}

We wish to determine the stresslet of the resulting flow, which can be written as
\begin{align}
    \stresslet &= \int_{S} %
    \left\{ 
    \frac{1}{2}[\Gammatensor] + \frac{1}{2}[\Gammatensor]^{T} - \frac{1}{3}\text{Tr}[\Gammatensor] \idmat \right\}
    \dd S,
\end{align}
and the tensor, $\Gammatensor$, can be written for an arbitrary slender looped swimmer as 
\begin{align}
    \Gammatensor &= \surf (\stress\cdot \no) - 2 \mu \bv{u} \no
    =\left[\bv{r}(\sdum) + \epsslend \erho(\sdum, \thetadum) \right] \bv{f}(\sdum,\thetadum) 
    -2 \mu \vslip(\sdum,\thetadum) \no(\sdum,\thetadum), 
\end{align}
where $\bv{u}$ is the fluid velocity and $\stress$ is the stress tensor, and we recall that the vector $\surf$ denotes a point on the surface of the swimmer (which should not be confused with the stresslet tensor $\stresslet$). 

Note that, since the uniform torus will not swim, the surface fluid velocity is equal to the slip velocity, $\bv{U}=\vslip$. For a uniform mobility, $\mobility$, Eq.~\eqref{Uo_uniformdonut} shows that the first mode of the surface fluid velocity is $\bvUo(s) = U_0 \norhat(s)$ with $U_0 =- \pi^2 \mobility \left[ \logfactor - 3/2 \right]$. From the symmetries of the Stokes flow equations, we conclude that the surface traction due to such a slip flow on a torus can only have a component in the normal direction and with a constant magnitude, $\bvfo(s) = f_0 \norhat(s)$. Using the slender body theory of~\citet{KoensLauga2018}, the relation between the traction $f_0$ and fluid velocity $U_0$ is given by 
\begin{align}
       f_0  &=   \frac{4U_0}{
       2\logfactor
       -5 
 }  + O(\epsslend) 
  =  -2 \pi^2 \mobility ~\frac{  
    2\logfactor
    - 3   }{
    2\logfactor
    -5 
 }  + O(\epsslend). 
\end{align}

Noting that the surface normal $\no = \erho$ and so the second term in the expression of $\Gammatensor$ integrates to zero since $\int \vslip\erho \dd\theta = \vslipcoeffcosmode{1}\norhat+\vslipcoeffsinmode{1}\binorhat = 0$, as well as $\vslip\cdot\erho=0$. Therefore, we can evaluate the surface integrals of $\Gammatensor$ as
\begin{align}
    &\int_{S} \Gammatensor ~\dd S 
    = \epsslend \int_{-1}^{1} \bv{r} \bvfo ~\dd s 
    = -\frac{\epsslend f_0}{\pi} \int_{-1}^{1} \norhat\norhat ~\dd s,
    \\
    &\int_{S} \text{Tr}[\Gammatensor] ~\dd S 
    = \epsslend \int_{-1}^{1} \bv{r} \cdot \bvfo ~\dd s  
    = -\frac{2\epsslend f_0}{\pi},
\end{align}
where we have used that $\bv{r}=-\norhat/\pi$ for a circular centreline of length 2. The integral $\int \norhat\norhat ~\dd s = \text{diag}(1,1,0)$.
Hence, the leading-order expression for the stresslet due to a uniformly-coated torus is
\begin{align}
    \stresslet 
    &=
    \frac{2 \pi \mobility \epsslend}{3}
    \frac{ \left[
    \log \left(  \frac{64 }{ \epsslend^2\pi^2 }  \right)     - 3  \right]}{\left[
    \log\left(\frac{64 }{\epsslend^2\pi^2}\right) -5 \right]
 }   \begin{pmatrix}
    1 & 0 & 0\\ 0& 1&0 \\ 0 & 0 & -2
    \end{pmatrix} +O(\epsslend^2).
    \end{align}

Rather remarkably, our slender theory has made it possible to derive the strength of this straining flow analytically for a uniformly-active torus. 
The stresslet of an active particle is 
important for understanding far-field interactions of active particles with boundaries and other particles, and is a key component in developing theory for active suspensions.

\section{Propulsion from geometric asymmetry}
 
\begin{figure}

    \centering
    \subfloat[Squashed torus]{\includegraphics[width=0.3\textwidth,viewport=300 300 860 700,clip]{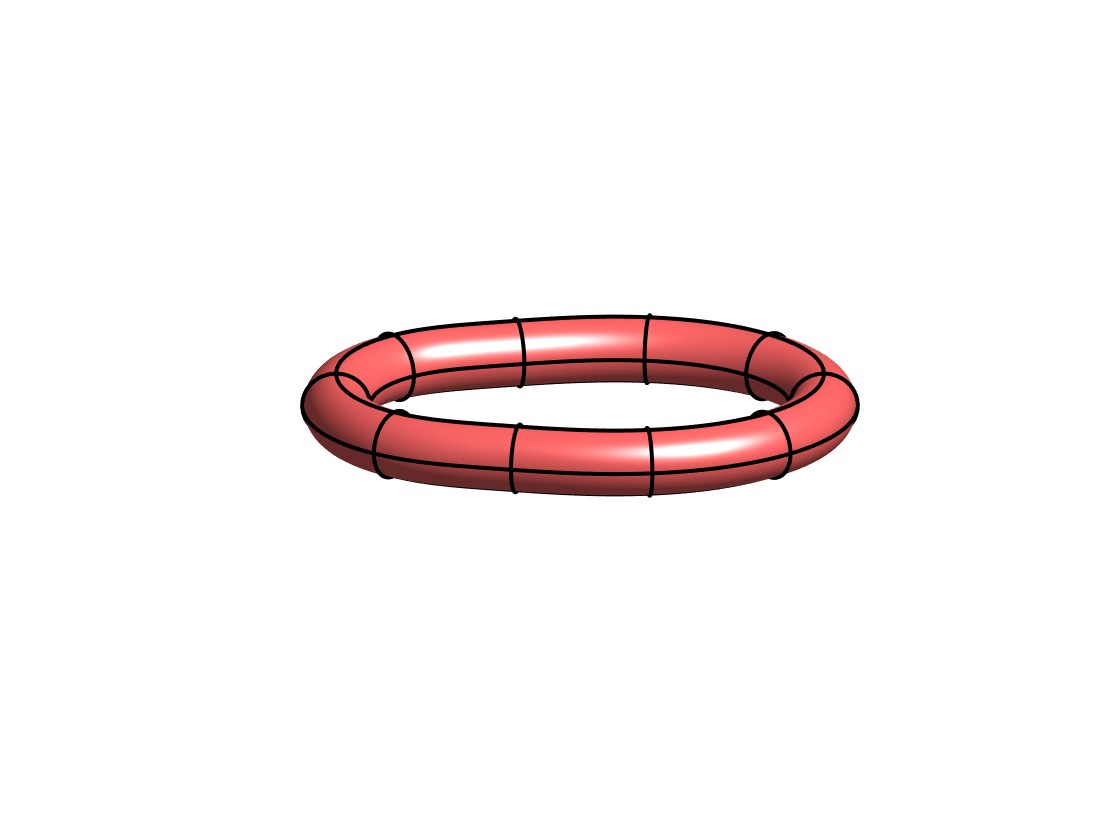}}
    \subfloat[Centreline profiles]{\includegraphics[width=0.25\textwidth,viewport=80 580 210 740,clip]{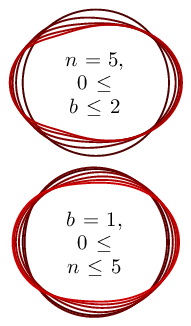}}
    \subfloat[Swimming speed]{\includegraphics[width=0.35\textwidth,viewport=50 570 250 750,clip]{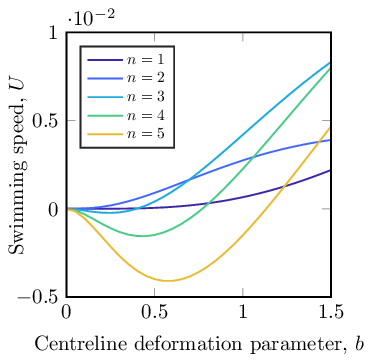}}
    \caption{Deforming the centreline of a uniform torus can induce swimming. 
    (a) A squashed torus that translates in-plane, which has a constant cross-sectional radius but asymmetric centreline. 
    (b) Profiles of the deformed centreline for the considered squashed tori with fixed $\deformcentrelineshape =5$, and varying $\deformcentrelineamp=0,0.5,1,1.5,2$. 
    (c) Centreline profiles when varying $\deformcentrelineshape$ from 0 to 5 with $\deformcentrelineamp=1$ fixed. Note that when $\deformcentrelineamp=0$ or $\deformcentrelineshape=0$, the centreline is a circle and the slender object is a uniform torus. 
    (d) In-plane swimming speed for the squashed torus as a function of the deformed centreline parameter, $\deformcentrelineamp$, for various fixed $\deformcentrelineshape=1,\dots5$ when $\epsslend=10^{-2}$. Positive $U$ corresponds to rightward swimming in (a). These squashed tori do not rotate.
    }
    \label{fig:SquashedTorus}
\end{figure}

While a uniform symmetric torus cannot swim, it is well-established that particles with uniform chemical activity may swim via geometric asymmetries, which lead to variations in solute concentration near the particle surface~\cite{michelin2015autophoretic, michelin2015geometric,michelin2017geometric}. 
We now use our theory to demonstrate that a squashed uniform torus may translate, and a bumpy uniform torus may rotate. 

We begin by maintaining a constant circular cross-section, and deforming the centreline of the torus in the manner given in cylindrical polar coordinates $(r,\varphi,z)$  by
\begin{equation}
    r = \frac{\alpha}{\pi}\left[1 + \deformcentrelineamp\cos^{2\deformcentrelineshape} \left(\frac{\varphi}{2} \right) \right],
\end{equation} 
where $\deformcentrelineshape$ is an integer, with higher $\deformcentrelineshape$ corresponding to a more localised deformation, and $\deformcentrelineamp$ is the amplitude of the centreline deformation. Here, $\alpha$ is chosen to rescale the geometry such that the arclength of the swimmer is 2. Note that $\varphi \in [0,2\pi)$ is not an arclength parameterisation, and so we must numerically find an appropriate representation of the curve for use in our Slender Phoretic Theory. 

An example of such a deformed torus, which we term a ``squashed" torus, is shown in Fig.~\ref{fig:SquashedTorus}a. In Fig.~\ref{fig:SquashedTorus}b, we show examples of the centreline profile when varying $\deformcentrelineamp$ and $\deformcentrelineshape$ independently. Note that when $\deformcentrelineamp=0$ or $\deformcentrelineshape=0$, the centreline is circular and the filament is the undeformed uniform torus previously considered in \S\ref{sec:UniformTorus}.

The swimming speed of these squashed slender tori are calculated numerically using our slender theory and plotted in Fig.~\ref{fig:SquashedTorus}c for different values of $\deformcentrelineshape$ when the slenderness $\epsslend=10^{-2}$. We see that these asymmetric centreline deformations can generate small, but significant, swimming speeds without rotation. The direction of motion depends sensitively on the precise shape, with possibility of motion in either direction along the symmetry axis as the deformation from a circular centreline is increased.

\begin{figure}
    \centering
    \subfloat[Bumpy torus]{\includegraphics[width=0.35\textwidth,viewport=300 300 860 700,clip]{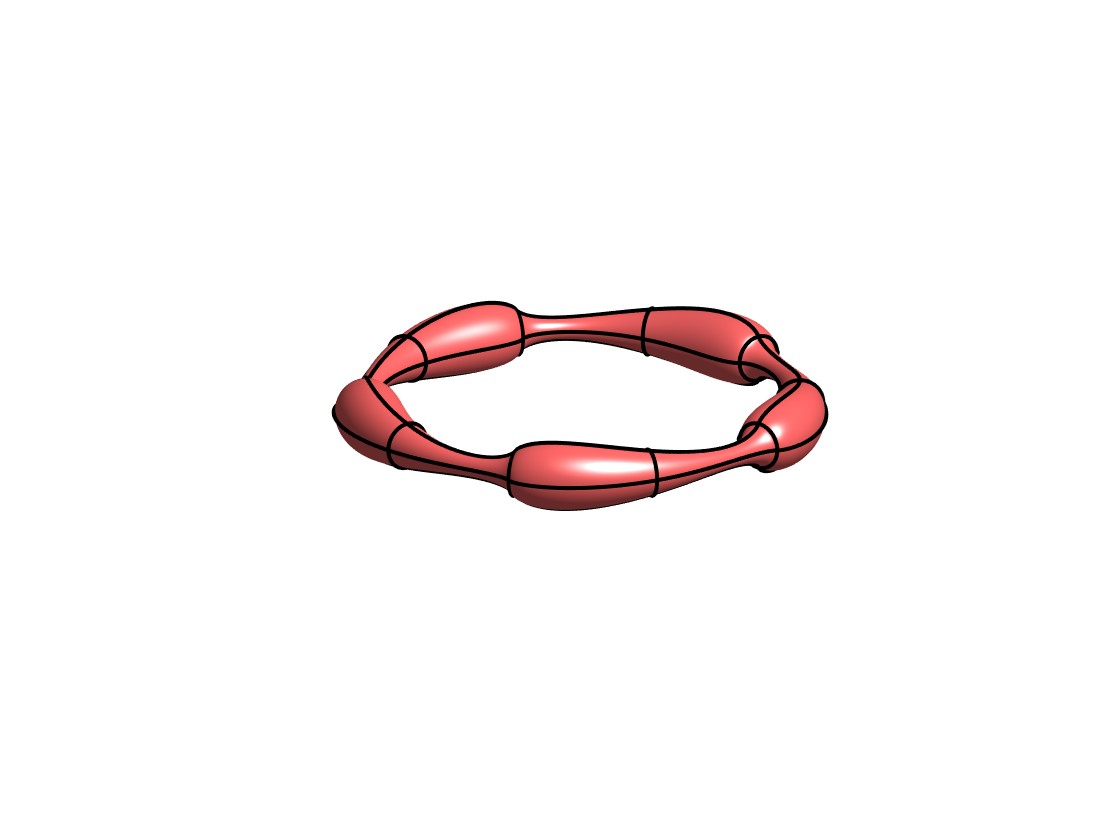}}
    %
    %
    \subfloat[Rotation speed]{\includegraphics[width=0.5\textwidth,viewport=50 550 300 750,clip]{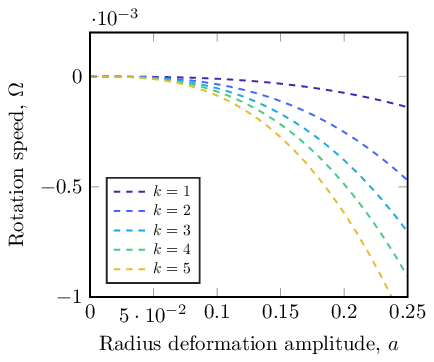}}
    \caption{Deforming the cross-sectional radius of a uniform torus can induce swimming. 
    (a) A bumpy torus that rotates, with a circular centreline but varying cross-sectional radius. 
    (b) Rotation speeds as a function of the bump amplitude, $\deformradiusamp$, for different numbers of bumps, $\deformradiusnum=1,2,3,4,5$, when $\epsslend=10^{-3}$. For $k\geq2$, these bumpy tori do not translate. The $k=1$ bumpy torus has a translational speed $U\approx0.75\deformradiusamp$ towards the raised bump. 
    }
    \label{fig:Bumpy torus}
\end{figure}
 
Alternatively, we can maintain a circular centreline, and generate asymmetry by varying the cross-sectional radius. By creating longitudinally asymmetric (sawtooth-esque) bumps, concentration gradients can be created in a similar manner to~\citet{michelin2015geometric}. If these bumps are arranged in a rotationally-symmetric manner, then we may expect this geometric alteration to lead to rotational motion. 
For a finite number of asymmetric bumps around the loop, we may expect the phoretic swimmer to rotate if these bumps all point in the same direction around the curve, whereas a net swimming motion may be achieved if not.

Inspired by previous work on pumping in autophoretic channels~\cite{michelin2015geometric},
we consider one possible deformation of the torus cross-sectional radius that retains some rotational symmetry, in the form of a sheared sine wave, for which we utilise the second Clausen function. The cross-sectional radius of the loop is given by
\begin{equation}
     \rho(s) = \beta\left[1 + \deformradiusamp\sum_{n=1}^{\infty}{\frac{\sin \deformradiusnum n\pi s}{n^2}}\right]
\end{equation}
where $\deformradiusamp < 1$  is the amplitude of the variation in cross-section and $\deformradiusnum$ gives the number of waves along the loop. The constant $\beta$ here is chosen to enforce that $\max\crossradius=1$. An example of this swimmer with $\deformradiusnum=5$ is shown in Fig.~\ref{fig:Bumpy torus}a. By symmetry, we might expect swimmers to rotate for $\deformradiusnum \geq 2$, while the case $\deformradiusnum = 1$ might result in a loopy drift motion. 

The resulting swimming for various fixed $\deformradiusnum$ and varying $\deformradiusamp$ is shown in Fig.~\ref{fig:Bumpy torus}b. For $\deformradiusnum\geq2$, the deformed torus spins about its central axis, with negligible translational velocity, as expected from symmetry arguments. Increasing the number of bumps increases the rotational speed. However, when there is only one bump on the torus ($\deformradiusnum=1$), these symmetry arguments do not apply, and the slender deformed torus translates in-plane with a smaller rotation about its central axis, which would combine to give a large looping trajectory in the plane of the torus.

\section{Twists and knots}

\begin{figure}
\centering
\subfloat[Twisted torus]{\includegraphics[width=0.3\textwidth,viewport=300 300 850 600,clip]{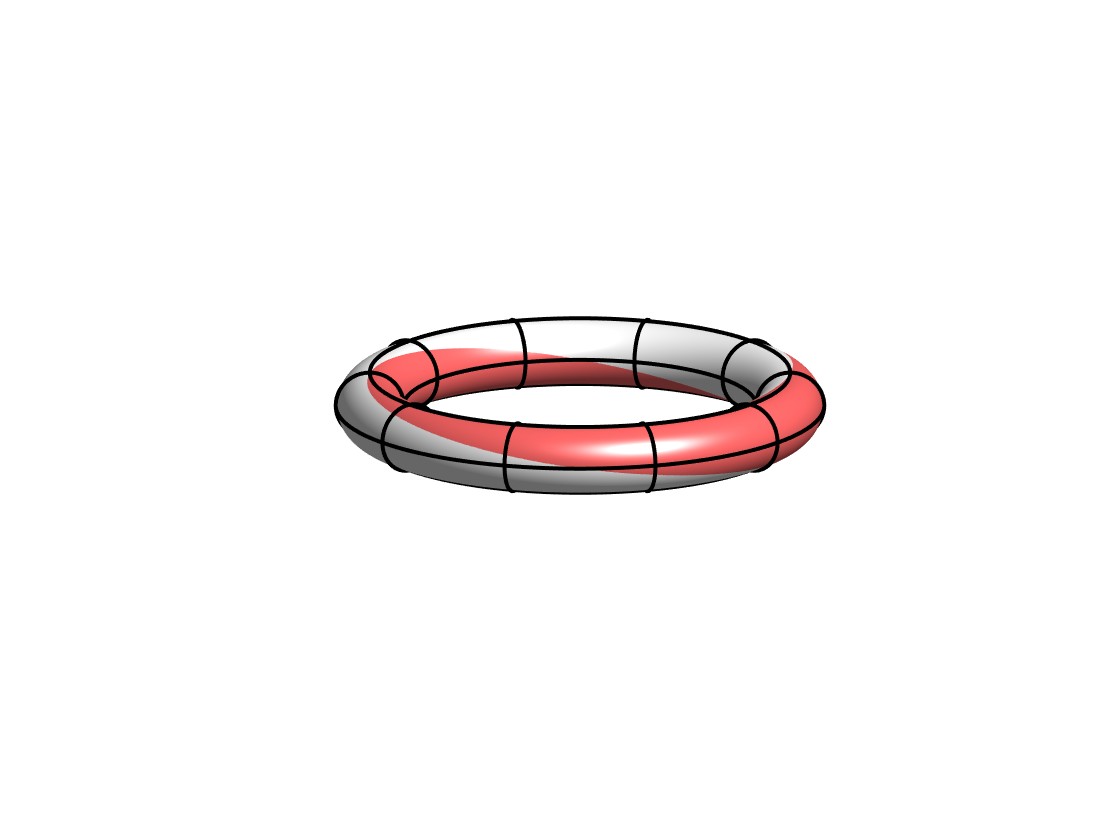}}
\hspace{0.5cm}
\subfloat[Glazed trefoil]{\includegraphics[width=0.3\textwidth,viewport=300 250 850 650,clip]{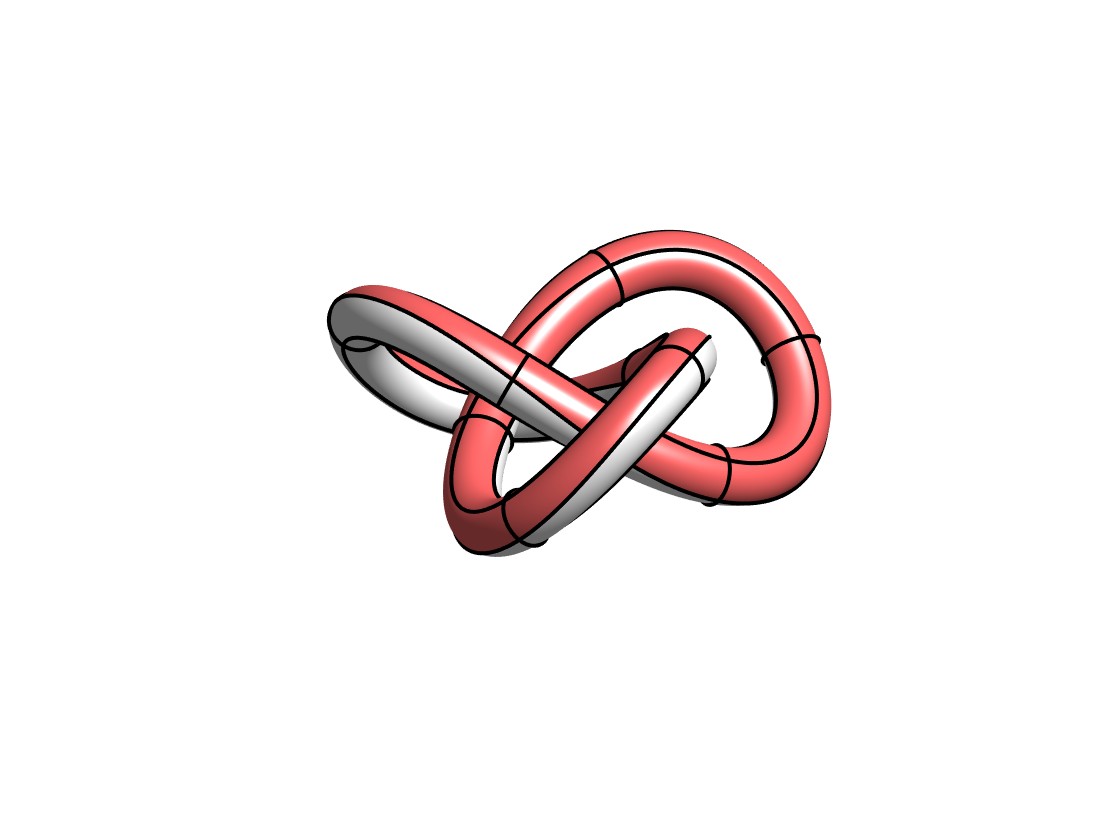}}
\hspace{0.5cm}
\subfloat[Dunked trefoil]{\includegraphics[width=0.3\textwidth,viewport=300 250 850 650,clip]{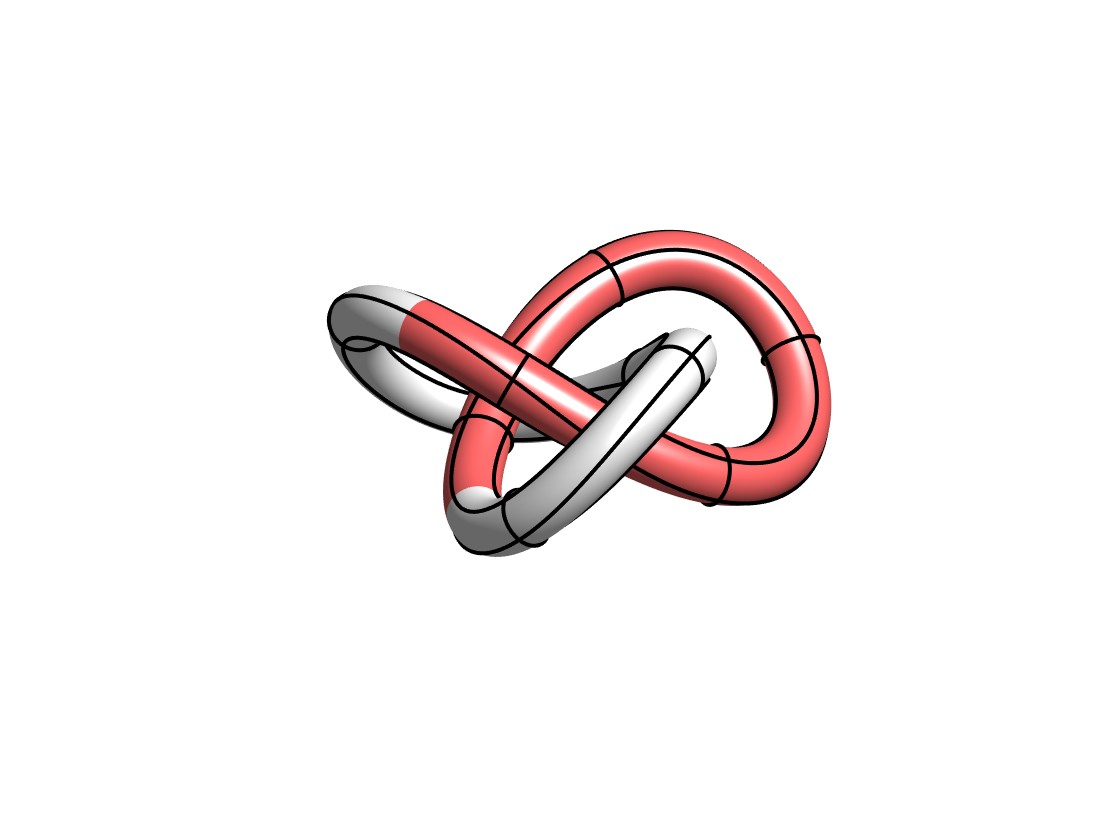}}
\caption{Geometric and topological variations from uniformly active tori can induce swimming. 
(a) A twisted torus, with an active strip that rotates around both the longitudinal and azimuthal directions, swims in-plane and rotates.
(b) A glazed trefoil, topologically distinct from a torus, swims and rotates about its rotational symmetry axis.
(c) A dunked trefoil swims in-plane and rotates about its swimming direction.
}
\label{fig:TwistKnotExamples}
\end{figure} 

We conclude our results with a demonstration of the flexibility of our theory to examine more complex surface chemistries and looped geometries, such as twists and knots.

\begin{figure}
    \centering

     \subfloat[Glazed swimming]{\includegraphics[width=0.45\textwidth,viewport=50 550 300 750,clip]{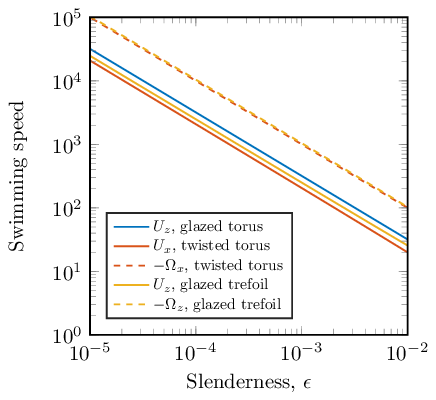}}
    \subfloat[Dunked swimming]{\includegraphics[width=0.45\textwidth,viewport=50 550 300 750,clip]{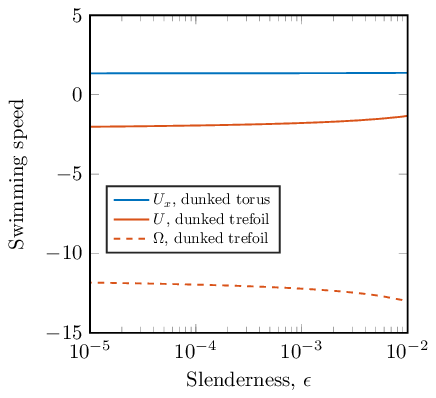}}
    
    \subfloat[Dunked torus]{\includegraphics[width=0.45\textwidth,viewport=50 550 300 750,clip]{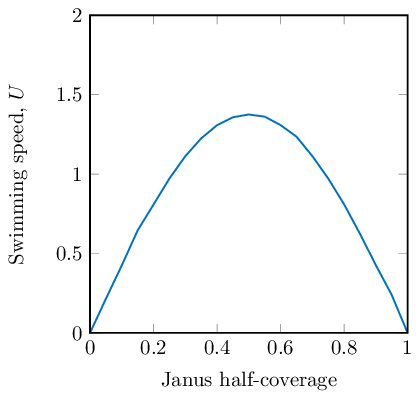}}
    \subfloat[Dunked trefoil]{\includegraphics[width=0.45\textwidth,viewport=50 550 300 750,clip]{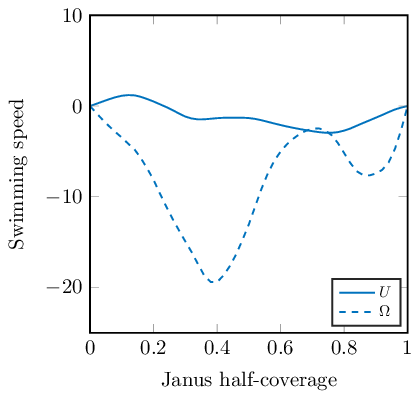}}
    
    \caption{Swimming behaviour of tori and trefoils with different chemical patterning. Solid lines represent translational velocities, dashed lines are rotational velocities. 
    (a) The glazed torus (blue), twisted torus (red) and glazed trefoil (yellow) swim with a dimensionless speed of $O(1/\epsslend)$. Both glazed examples swim parallel to the axis of rotational symmetry, $\bv{e}_z$, whereas the twisted torus swims in-plane. The twisted torus and glazed trefoil rotate (clockwise) about their swimming axes, with rotational velocities of $O(1/\epsslend)$. 
    (b) The dunked torus and dunked trefoil swim in-plane, with the dimensionless swimming speed appearing near-constant with $\epsslend$. The dunked trefoil also rotates about its swimming axis.
    (c) The (in-plane) swimming speed for a partially-dunked torus with activity $\activity=-1$ in $s\in(-s_c,+s_c)$ and $\activity=0$ elsewhere. Here the slenderness is $\epsslend=10^{-2}$, and there is no rotation for any coverage. 
    (d) The in-plane swimming and rotation speed of a dunked trefoil as a function of the half-coverage, when the activity is centred around the extremity of a loop (such as the example shown in Fig.~\ref{fig:TwistKnotExamples}c, corresponding to a half-coverage 0.5).
    }
    \label{fig:MainResults}
\end{figure}

We begin by ``twisting'' the activity strip of a glazed torus as it passes around the loop (Fig.~\ref{fig:TwistKnotExamples}a). The resulting dimensionless swimming behaviour is plotted in Fig.~\ref{fig:MainResults}a as a function of the slenderness. The translational swimming speed is comparable to that of the glazed torus, with a speed $U\approx0.21/\epsslend$. However, while the regular glazed torus swims out-of-plane with no rotation, this twisted torus swims in-plane while rotating about its swimming axis, with a rotation speed $\Omega\approx1.00/\epsslend$.

We are also able to determine the swimming behaviours for more complex topologies than the torus, for example knots. Here, we focus on a regular representation of the trefoil,  given by the parameterisation
\begin{align}
     x =& \gamma[\sin\phi + 2\sin2\phi], 
     \nonumber \\ 
     y =& \gamma[\cos\phi - 2\cos2\phi], 
     \\
     z =& -\gamma\sin3\phi,
     \nonumber
\end{align}
where $\phi \in [0,2\pi),$ and $\gamma$ is a constant chosen such that the total arclength is equal to 2.
The considered examples are shown in Fig.~\ref{fig:TwistKnotExamples}b \& \ref{fig:TwistKnotExamples}c. As with the tori, we consider ``glazed" and ``dunked" trefoils, which are half active and half inert. 

The glazed trefoil, with activity variation around its slender cross-section, translates similarly to the glazed torus, with a speed $U\approx0.25/\epsslend$ in the out-of-plane direction, $\bv{e}_z$, as shown in Fig.~\ref{fig:MainResults}a. However, due to the chirality of the trefoil, it also rotates about its swimming axis, with a rotational speed $\Omega\approx1.04/\epsslend$. 

Similarly, the dunked trefoil's swimming behaviour is much like the dunked torus, but with added rotation about the swimming direction, as shown in Fig.~\ref{fig:MainResults}b. Both the dimensionless swimming and rotation speeds remain approximately constant with the slenderness, $\epsslend$, but, perhaps surprisingly, the dunked trefoil translates in the opposite direction to the torus, away from the active end, with a significantly larger rotation speed. 

This barrel-rolling swimming behaviour of the dunked trefoil has a complex dependence on the coverage of the chemical activity, with swimming possible in either direction depending on how much of the knot is coated. In Fig.~\ref{fig:MainResults}d, we show the numerically calculated swimming velocities of a dunked trefoil as the coverage of the active region is varied. In particular, it shows the translational and rotational speed for an active region that extends for an arclength $s_c$ either side of the one of the most extreme points of the loops. We observe a complex relation between coverage and swimming speed that is neither symmetric nor monotonic. In fact, the direction of swimming changes as the coverage varies. The maximum swimming observed is approximately $U\approx-3$ when $s_c\approx0.75$, whereas the maximum rotation occurs for a different activity, with $\Omega\approx-19$ near $s_c\approx0.4$. This non-trivial behaviour occurs as the activity interface moves around the knot, causing the strong slip flows close to the jump in activity to be oriented in different directions, as well as variation in the contributions to the surface concentration due to  points that are distant in arclength but relatively close in space (but still far relative to the slenderness).

\section{Outlook}

Slender and flexible active filaments are an exciting new class of self-propelled active matter, with a rich variety of dynamical behaviours. In this work, we have studied the swimming of such looped and knotted active filaments.
The design space of possible filamentous knots grows staggeringly quickly with the number of possible crossings, and that is before variations in surface chemistry and cross-sectional radius are incorporated. We have provided a rapid and simple theoretical approach, that amounts to evaluating a line integral and its derivatives, followed by solving a line integral equation, to elucidate the dynamics of these active filaments. 

This complements our recent work on a slender theory for unlooped active filaments~\cite{katsamba2020slender,katsamba2022chemically}, with the important distinction that, in the looped case, there are no ``end-point'' effects, where the asymptotic assumption of slenderness breaks down, making the slender body theory accurate at every point on the filament. 

Crucially, the theory is amenable to closed form solutions for the dynamics, particularly in the case of the torus, for which we were able to calculate swimming speeds and pumping strengths analytically. These solutions can not only provide a benchmark for testing other numerical schemes, but also generate simple expressions to be used as a basis for suspension modelling. Furthermore, the speed of calculation makes the theory suitable for coupling to elastic beam/loop equations, allowing rapid simulation of flexible phoretic loops. Our (currently unoptimised) code~\cite{SPTcode} typically solves within seconds in MATLAB on a laptop, comparable to state-of-the-art, highly-optimised routines for non-asymptotic boundary integral representations running on multiple cores~\cite{malhotra2024efficient}, yet has room for significant improvements in both accuracy and runtime.

Additionally, the framework of our theory may be adapted to account for more complicated reactions than the zeroth order kinetics that we have presented. For example, in first order kinetics (finite Damk\"ohler number), which may be required to reproduce the size-dependent swimming speeds observed experimentally \citep{ebbens2012size}, the activity is proportional to the local solute concentration, and so we may expect the equivalent of our Slender Phoretic Theory to result in (implicit) integral equations for the surface concentration, that could then be solved numerically. Increases in Damk\"ohler number typically tend to decrease swimming speeds, but without significant qualitative impact on dynamics \cite{yariv2019self}.

From a small number of geometries and chemical patterns, we have found several distinct fluid flows and modes of swimming, including: perpendicular and parallel translation, bulk pumping with and without rotation, stationary spinning, circling and barrel rolling. We hope that, in the near future, our theory can be utilised to explore the full extent of parameter space, and impart novel functionality to active matter systems. 

One potential area of interest is in active colloids research, where slender active filaments may have rich new dynamical behaviours, as seen in related systems with adaptive locomotion through constrictions~\cite{HuangAdaptiveSciAdv}, dynamic clustering of Brownian rings \cite{theeyancheri2024dynamic} and spontaneous oscillations~\cite{laskar2017filament}. In nature, colloids (such as in milk, smoke, fog) are generally topological spheres, as their shape is driven by minimising interfacial tension. However, recent work on these passive (non-swimming) colloids has shown that particle topology can drive interesting phenomena. For instance, when multiply-connected particles were introduced into a nematic liquid crystal, the colloids created topological defects in the liquid crystal that were dictated by the particle topology~\cite{senyuk2013topological}. Similarly, knotted colloids fabricated via two photon polymerisation (2PP) were able to pattern bulk and surface defects in nematic fluids~\cite{martinez2014mutually}. These results in passive topological colloids suggest that tying the knot on active filaments may provide further behaviours, particularly in bulk, that we have yet to predict and are only beginning to be explored \cite{delmotte2024scalable}, and we hope that understanding the behaviour of single particles, as discussed here, is the first step towards achieving this.

\section*{Acknowledgements}

P. K. was supported by the project ``SimEA'', funded by the European Union’s Horizon 2020 research and innovation programme under Grant Agreement No. 810660, and in part by the Engineering and Physical Sciences Research Council (EPSRC) grant EP/R041555/1 ‘‘Artificial Transforming Swimmers for Precision Microfluidics Tasks’’ to T. M.-J. 
M. B. was supported by a Clifford Fellowship at University College London, and the Leverhulme Trust Research Leadership Award ‘‘Shape-Transforming Active Microfluidics’’ to T. M.-J. 
L. K. was funded by Australian Research Council (ARC) under the Discovery Early Career Research Award scheme (grant agreement DE200100168).

\appendix

\section{Derivation of the Slender Phoretic Theory of slender active loops}  \label{App_derivation}

Here we give a full derivation of the Slender Phoretic Theory for looped filaments. 
 
\subsection{Boundary Integral Equation for the diffusion equation}

We use Green's second identity for the concentration field $c(\xo)$ and the well-known 
Green's function, $ \Greendiff (\x,\xo)= \frac{1}{4\pi |\x-\xo|}$, for Laplace's equation in an unconfined, three-dimensional region, forced by a point sink at $\xo$. This Green's function has $\nabla^2\Greendiff(\x,\xo)=-\delta(\x-\xo)$. (Note the translational invariance in our unbounded domain, hence the notation $\Greendiff(\x,\xo) = \Greendiff(\x-\xo) = 1/4\pi|\R|$, where $\R=\x-\xo$.)
As in~\citet{katsamba2020slender}, we apply Green's second identity to the fluid domain outside the filament, $V$, bounded by the filament surface $S$ with normal $\no$ pointing out of the filament, to obtain
\begin{equation}
\lambdaf c(\x)=\frac{1}{4\pi}\int_{S}^{}\left[ \frac{c(\xo) \no \cdot\left(\x-\xo\right)}{ |\x-\xo|^3} + \frac{\activity(\xo)}{|\x-\xo|}\right] \dd S (\xo) ,\label{diffusionBIderived}
\end{equation}
where $\lambda=1/2$ for $\x\in S$, $\lambda=1$ for $\x\in V$, and $\lambda=0$ for $\x\notin V$~\citep{Pozrikidis1992}. 

Substituting the filament geometry into Eq.~\eqref{diffusionBIderived} for two points on the filament surface, $\x=\surf(s,\theta)$ and $\xo=\surf(\sdum,\thetadum)$, and setting $\lambdaf=1/2$, gives the boundary integral formula 
\begin{align} 
2\pi c(s,\theta)=
&  \int\limits_{-1}^{1} \!\int\limits_{-\pi}^{\pi}
\left[
\frac{\activity(\sdum,\thetadum)}{  |\bvR|} \bigg|\ddsprime{\surf}\times\ddthprime{\surf}\bigg| 
\right. 
\left.
- \frac{c(\sdum,\thetadum)  \bvR}{ |\bvR|^3} \cdot\left(\ddsprime{\surf}\times\ddthprime{\surf}\right)
\right]\, 
\dd \thetadum ~\dd \sdum , \label{diffusionBIgeneral}
\end{align}
where the vector between two points on the surface is
\begin{align}
	\bvR &\equiv\surf(s,\theta)-\surf(\sdum,\thetadum)  
	=\bvRo(s,\sdum) + \epsslend\left[\crossradius(s)\erho(s,\theta) - \crossradius(\sdum)\erho(\sdum,\thetadum)\right], \label{bvRgeneral}
\end{align}
and $\bvRo\equiv \bv{r}(s) - \bv{r}(\sdum)$ is a vector joining the points on the centreline.
Note that the surface element and its magnitude are given by
\begin{align}
 \dds{\surf}\times\ddth{\surf} 
	&=\epsslend \crossradius(s)  \left( \epsslend \ddrhods \tanhat(s)  
	 -\left[1-\epsslend\crossradius(s) \curvature(s) \cos\dtheta\right]
	 \erho(s,\theta)    \right),\label{surfelementvector} \\
\bigg|\dds{\surf}\times\ddth{\surf}\bigg| &= \epsslend \crossradius(s)  \sqrt{ \left[\epsslend \ddrhods\right]^2 
+\left[1-\epsslend\crossradius(s) \curvature(s) \cos\dtheta\right]^2
}.	 \label{surfelementscalar}
\end{align}

Similarly to~\citet{KoensLauga2017_AnalytSolSRT}, in the case of looped filaments we replace $\sdum$  by $s + q$,  and $\int_{-1}^{1}\dd \sdum$ by $\int_{-1}^{1} \dd q$, so that
 \begin{align}
	2\pi c(s,\theta) = 
	\int\limits_{-1}^{1}\!\int\limits_{-\pi}^{\pi} 
	\left[\kerone(s,\theta, s + q,\thetadum) + \kertwo(s,\theta,s + q,\thetadum)\right] 
	\dd\thetadum\dd q, 
\end{align}
where the expanded kernels $\kerone(s,s + q,\theta,\thetadum), \kertwo(s,s + q,\theta,\thetadum)$ are the same as in~\citet{katsamba2020slender}, but with $\sdum$ replaced by $s + q$, so that 
\begin{subequations}\label{kernels}
\begin{align}
\kerone&=	~\frac{\crossradius(s + q)\activity(s + q,\thetadum)}{ |\bvR|}    \quad\left[ 
1 - \epsslend\crossradius(s + q)\curvature(s + q)\cos \Theta(s + q, \thetadum)
+ O(\epsslend^2)\right], \label{keroneref} \\
\kertwo&=\epsslend\frac{\crossradius(s + q)c(s + q,\thetadum) }{ |\bvR|^3}\bvR\cdot   \bigg\{
\erho(s + q,\thetadum) 
\nonumber\\
&\qquad\qquad
- \epsslend \left[ \crossradius'(s + q)\tanhat(s + q) +\crossradius(s + q) \curvature(s + q) \cos\Theta(s + q,\thetadum)  \erho (s + q,\thetadum)\right]
\bigg\}. \label{kertwodef}
\end{align}
\end{subequations}

\subsection{Matched asymptotic expansion}

\subsubsection{Outer region}

In the outer region, points are relatively far apart with $q=O(1)$, hence points on the centreline are separated by $\bvRo(s,s+q)=O(1)$, and so we can approximate the vector between the two points on the surface by the difference between the centreline near them, $\bvR \approx\bvRo$. 
Expanding in powers of $\epsslend$ to first order, we find that
\begin{align}
\bvR=\bvRo(s,s+q) + \epsslend \Deltaerhorhoq,  \quad
    \Deltaerhorhoq\equiv \crossradius(s)\erho(s,\theta)- \crossradius(s+q)\erho(s+q,\thetadum).  
 \end{align} 
 
The outer expansions, $\outerexp{\kerone}$ and $\outerexp{\kertwo}$, of the kernels $\kerone$ and $\kertwo$, as defined in Eqs.~\eqref{keroneref}-\eqref{kertwodef}, are similarly expanded to give
\begin{align}
&  \outerexp{\kerone}   
=	~\frac{\crossradius(s+q)\activity(s+q,\thetadum)}{ |\bvRo(s,s+q)|}    
\nonumber\\
& \qquad\qquad
\times 
\left\{1-\epsslend \left[\crossradius(s+q)\curvature(s+q)\cos\Theta(s+q,\thetadum)+\frac{\bvRo}{|\bvRo|^2}\cdot\Deltaerhorhoq \right]  +  O(\epsslend^2)\right\}
\label{K1_outer},\\
&\outerexp{\kertwo}  =\epsslend\frac{\crossradius(s+q)c(s+q,\thetadum)}{ |\bvRo|^3}
\Bigg\{
\bvRo\cdot\erho(s+q,\thetadum) + \epsslend \left(\Deltaerhorhoq\right)\cdot\erho(s+q,\thetadum)\nonumber \\
&\qquad\qquad\qquad
- \epsslend \bvRo\cdot\left[ \crossradius'(s+q)\tanhat(s+q) +\crossradius(s+q) \curvature(s+q) \cos\Theta(s+q,\thetadum) \erho (s+q,\thetadum)\right]  \nonumber \\
&\qquad\qquad\qquad\qquad\qquad\qquad\qquad\qquad
- \epsslend (\bvRo\cdot\erho(s+q,\thetadum)) \frac{3\bvRo\cdot\Deltaerhorhoq}{|\bvRo|^2} 
+O(\epsslend^2) 
\Bigg\}.
    \label{K2_outer}
\end{align}


\subsubsection{Inner region expansion} 

In the inner region, the distance between points is small so that $q=O(\epsslend)$  
and we let $q=\epsslend\chi$, where $\chi$ is O(1). The displacement vector, $\bvR$, can then be expanded as
\begin{equation}
    \bvR = \epsslend \bvRione + \epsslend^2\bvRitwo  + O(\epsslend^3),
\end{equation}
where
\begin{align}
    \bvRione &= 
    -\chi \tanhat(s) 
    + \crossradius(s)[\erho(s,\theta)-\erho(s,\thetadum)], 
    \\
    \bvRitwo &= 
    - \left[ \frac{\chi^2}{2}\curvature(s) \norhat(s) 
    + \chi \ddrhods \erho(s,\thetadum) 
    - \chi\crossradius(s)\curvature(s)\cos\dtheta(s,\thetadum)\tanhat(s) 
    \right].
\end{align}

The inner kernels, $\innerexp{\kerone}$ and $\innerexp{\kertwo}$ are then found to be 
\begin{align}
\innerexp{\kerone}&=
\frac{\crossradius(s)\activity(s,\thetadum)}{ \epsslend\sqrt{\chi^2+\gamma^2}}
- \frac{\activity(s,\thetadum)}{\sqrt{\chi^2+\gamma^2}}
\crossradius^2(s)\curvature(s)\cos\dtheta(s,\thetadum)
\nonumber\\
&\quad
+\frac{\crossradius^2(s)\activity(s,\thetadum)}{(\chi^2+\gamma^2)^{3/2}}
\bigg[ \frac{\curvature(s)\chi^2}{2} \left[\cos\dtheta(s,\thetadum) + \cos\dtheta(s,\theta)\right] 
+ \chi \ddrhods [\cos(\theta-\thetadum)-1]\bigg]
\nonumber\\
&\quad
+ \frac{\chi }{\sqrt{\chi^2+\gamma^2}}
\partial_s\left[\crossradius(s)\activity(s,\thetadum) \right] 
+ O(\epsslend),
\end{align}
\begin{align}
\innerexp{\kertwo}
=&
\frac{\crossradius^2(s)c(s,\thetadum)  }{ \epsslend(\chi^2+\gamma^2)^{3/2}} \left[\cos(\theta-\thetadum)-1\right] 
+\frac{\chi \crossradius(s)}{ (\chi^2+\gamma^2)^{3/2}}   \left[\cos(\theta-\thetadum)-1\right]\partial_s\left[\crossradius(s) c(s,\thetadum)\right]\nonumber\\
&
-\frac{\crossradius(s)c(s,\thetadum)\curvature(s)\cos\dtheta(s,\thetadum)}{(\chi^2+\gamma^2)^{3/2}} 
\left[-\frac{\chi^2}{2}
+\crossradius^2(s)  \left[\cos(\theta-\thetadum)-1\right]\right]
\nonumber\\
&+3\frac{\crossradius^3(s)c(s,\thetadum)}{(\chi^2+\gamma^2)^{5/2}}
\left[\cos(\theta-\thetadum)-1\right]
\bigg\{\chi\ddrhods \left[\cos(\theta-\thetadum)-1\right]\nonumber\\
&\qquad\qquad\qquad\qquad\qquad\qquad\qquad\qquad
+\frac{\chi^2 }{2}\curvature(s) \left[\cos\dtheta(s,\thetadum) + \cos\dtheta(s,\theta)\right] 
\bigg\} \nonumber\\
&+O\left( \epsslend c\right), 
\end{align}
where we have used that
\begin{align}
|\bvRione|&=\sqrt{\chi^2+\gamma^2}, \\
\gamma^2&=2\crossradius^2(s)\left[1-\cos(\theta-\thetadum)\right].
\end{align}

We will perform the integration with respect to $q$, following~\citet{KoensLauga2017_AnalytSolSRT}, by using the   values of integrals of the form
\begin{align}
H^{i}_{j}&=\int_{-1}^{1}\frac{\chi^i}{\left(\chi^2+\gamma^2\right)^{j/2}}\dd q 
%
= \epsslend
\int_{\asinh{(-1/\epsslend \gamma)}}^{\asinh{(1/\epsslend \gamma)}} \gamma^{i-j+1} \frac{\sinh^i\phi}{\cosh^{j-1}\phi}\dd \phi,
\end{align}
where $\gamma$ is independent of $q$, $i$ and $j$ are positive integers, and we recall $q=\epsslend\chi$. To obtain the last integral here, we used the substitution $\chi = \gamma \sinh{\phi}$ (hence $\dd q = \epsslend\dd \chi = \epsslend \gamma \cosh \phi ~\dd \phi $), as in~\citet{KoensLauga2018}.

We note that $H^{i}_{j}=0$ when $i$ is odd, because then the integrand is an odd function of $q$ or $\phi$. The other relevant values of $H$ are calculated to be
\begin{equation}
\begin{aligned}
    &H^{0}_{1} = 2 \epsslend\log \left(\frac{2}{\epsslend\gamma}\right) + O(\epsslend^3\gamma), 
\qquad
  H^{0}_{3} = 
\frac{2 \epsslend}{\gamma^2} + O(\epsslend^3),
\\
 H^{2}_{3} &= \epsslend\left[ \log\left(\frac{4}{\epsslend^2\gamma^2} \right) - 2 \right] + O(\epsslend^3 \gamma^2),
 \qquad
  H^{2}_{5} = \frac{2\epsslend}{3\gamma^2} + O(\epsslend^3).
\end{aligned}
\end{equation}

Integrating the inner kernels with respect to $q$ therefore gives
\begin{multline}
\int_{-1}^{1} \innerexp{\kerone} \dd q=
\crossradius(s)\activity(s,\thetadum)
\Bigg\{
\left[
1  
-  
\epsslend \crossradius(s)\curvature(s)\cos\dtheta(s,\thetadum)\right]\left[   \log \left(\frac{4}{\epsslend^2\gamma^2}\right)  \right] 
\\
%
+ \frac{1}{2}\epsslend 
\crossradius(s)
 \curvature(s)   \left[\cos\dtheta(s,\thetadum) + \cos\dtheta(s,\theta)\right]    \log\left(\frac{4}{\epsslend^2 \gamma^2 e^2} \right)   \Bigg\}
+ O(\epsslend^2),
\end{multline}
%
%
%
\begin{multline}
 \int_{-1}^{1} \innerexp{\kertwo} \dd q
=
c(s,\thetadum)\Bigg\{
 -    1 
+ \epsslend\crossradius(s) \curvature(s)\cos\dtheta(s,\thetadum) 
    \left[ 1+\frac{1}{2}\log\left(\frac{4}{\epsslend^2 \gamma^2 e^2} \right)      \right]
\\
-\frac{1}{2}\epsslend\crossradius (s)\curvature(s) 
 \left[\cos\dtheta(s,\thetadum) + \cos\dtheta(s,\theta)\right]     \Bigg\}
+O\left( \epsslend^2 c\right).
%
\end{multline}

Expanding the logarithm term, we note that
\begin{align}
\log\left(\frac{4}{\epsslend^2\gamma^2}\right) 
&= \log 
\left( 
\frac{2}{\epsslend^2  \crossradius^2(s)\left[1-\cos(\theta-\thetadum)\right]}\right)
=
\log 
\left( 
\frac{2}{\epsslend^2  \crossradius^2(s)}\right) - \log\left[1-\cos(\theta-\thetadum)\right].
\label{eq:LogExpansion}
\end{align}

\subsubsection{Matching region}

To match the inner and outer solutions, we must first expand the outer solutions in terms of the inner variables (or vice versa, which gives the same result), for which we use the superscript $(\text{out})\in(\text{in})$. We find that the expanded outer kernels are
\begin{align}
\expouterkernelininner{\kerone}&=\frac{\crossradius(s)\activity(s,\thetadum)}{|q|}+\sign{q}\partial_s\left[\crossradius(s)\activity(s,\thetadum)\right]
\nonumber\\
&\qquad
  -\frac{1}{2}\epsslend\frac{\crossradius(s)\activity(s,\thetadum) }{|q|}\crossradius(s)\curvature(s)\left[
\cos\dtheta(s,\thetadum)-\cos\dtheta(s,\theta)
\right] 
 + O(\epsslend^2), \\
\expouterkernelininner{\kertwo}&=\epsslend^2\frac{\crossradius^2(s)c(s,\thetadum)}{ |q|^3}
\left[\cos(\theta-\thetadum)-1\right]\left[1+O(\epsslend
)\right].
 \label{K2_expouterkernelininner}
\end{align}

After integrating over $q$, we lose the $\sign{q}$ term so that the integrated kernals in the matching region are 
\begin{align}
\int_{-1}^{1}\dd q \int_{-\pi}^{\pi} \dd \thetadum ~\expouterkernelininner{\kerone}
&=\crossradius(s)\int_{-1}^{1}\frac{\dd q}{|q|}
\int_{-\pi}^{\pi} \activity(s,\thetadum)   \dd \thetadum 
\\
   &-\frac{1}{2}\epsslend\crossradius^2(s)\curvature(s)\int_{-1}^{1} \frac{\dd q}{|q|}\int_{-\pi}^{\pi}   \activity(s,\thetadum)  \left[
\cos\dtheta(s,\thetadum)-\cos\dtheta(s,\theta)
\right] \dd \thetadum 
 + O(\epsslend^2), \nonumber \\
\int_{-1}^{1}\dd q  \int_{-\pi}^{\pi} \dd \thetadum 
~\expouterkernelininner{\kertwo}
&=\epsslend^2\int_{-1}^{1}\frac{\dd q}{|q|^3} 
\int_{-\pi}^{\pi}  
\crossradius^2(s)c(s,\thetadum)
\left[\cos(\theta-\thetadum)-1\right] 
\left[1+O(\epsslend)\right]  \dd \thetadum
= O(\epsslend^2c).
\end{align}


\subsubsection{Full expression for the expansion of the concentration field} \label{full_sum_appendix}
The full boundary integral equations are approximated by the adding the outer and inner expansions and subtracting from each the common part, 
\begin{align}
&	2\pi c(s,\theta) \approx 
	\int\limits_{-1}^{1}\!\int\limits_{-\pi}^{\pi} 
	\left(
     \innerexp{\kerone} 
    + \innerexp{\kertwo}
    +\outerexp{\kerone}
    + \outerexp{\kertwo} 
    - \expinnerkernelinouter{\kerone}  
    - \expinnerkernelinouter{\kertwo} \right)
	\dd\thetadum\dd q\nonumber\\
%
%
%
	=&
	\int_{-\pi}^{\pi}
\crossradius(s)\activity(s,\thetadum)
\Bigg\{
\log\left(\frac{4}{\epsslend^2 \gamma^2} \right)  
+\epsslend \crossradius(s) \curvature(s) \Bigg[
\cos\dtheta(s,\theta) \log\left(\frac{2}{\epsslend \gamma e}\right)  
+\cos\dtheta(s,\thetadum)  \log\left( \frac{\epsslend \gamma}{2 e}\right)  
\Bigg]
  \Bigg\} \dd \thetadum
\nonumber
\\ 
%
%
%
&+\int_{-\pi}^{\pi}
c(s,\thetadum)\Bigg\{
 -    1 -\frac{1}{2}\epsslend\crossradius (s)\curvature(s) 
  \cos\dtheta(s,\theta)
+ \frac{1}{2}\epsslend\crossradius(s) \curvature(s)\cos\dtheta(s,\thetadum) 
    \left[ 1+\log\left(\frac{4}{\epsslend^2 \gamma^2 e^2} \right)      \right]
    \Bigg\} \dd \thetadum
\nonumber
  \\
&-\crossradius(s)\int_{-1}^{1}\frac{\dd q}{|q|}
\int_{-\pi}^{\pi} \activity(s,\thetadum)   \dd \thetadum 
   +\frac{1}{2}\epsslend\crossradius^2(s)\curvature(s)\int_{-1}^{1} \frac{\dd q}{|q|}\int_{-\pi}^{\pi}   \activity(s,\thetadum)  \left[
\cos\dtheta(s,\thetadum)-\cos\dtheta(s,\theta)
\right] \dd \thetadum 
\nonumber
\\
&+	~\int\limits_{-1}^{1}\!\int\limits_{-\pi}^{\pi}\frac{\crossradius(s+q)\activity(s+q,\thetadum)}{ |\bvRo(s,s+q)|}    
\left\{1-\epsslend \left[\crossradius(s+q)\curvature(s+q)\cos\Theta(s+q,\thetadum)+\frac{\bvRo}{|\bvRo|^2}\cdot\Deltaerhorhoq \right] \right\} \dd \thetadum \dd q
\nonumber
\\
&+\epsslend\int\limits_{-1}^{1}\!\int\limits_{-\pi}^{\pi}\frac{\crossradius(s+q)c(s+q,\thetadum)}{ |\bvRo|^3}
\Bigg\{
\bvRo\cdot\erho(s+q,\thetadum) + \epsslend \left(\Deltaerhorhoq\right)\cdot\erho(s+q,\thetadum)\nonumber \\
&\qquad\qquad\qquad\qquad\quad
- \epsslend \bvRo\cdot\left[ \crossradius'(s+q)\tanhat(s+q) +\crossradius(s+q) \curvature(s+q) \cos\Theta(s+q,\thetadum) \erho (s+q,\thetadum)\right]  \nonumber 
\nonumber
\\
&\qquad\qquad\qquad\qquad\qquad\qquad\qquad
- \epsslend (\bvRo\cdot\erho(s+q,\thetadum)) \frac{3\bvRo\cdot\Deltaerhorhoq}{|\bvRo|^2} 
\Bigg\} \dd \thetadum \dd q 
\nonumber
\\
&+O(\epsslend^2)  .
\label{eq:FullExpansion}
	\end{align}

\subsubsection{Leading-order concentration}

Removing terms of $O(\epsslend)$ or higher, we find an expression for the leading-order concentration field
\begin{align}
	2\pi c(s,\theta) \approx &
	\int_{-\pi}^{\pi}
\crossradius(s)\activity(s,\thetadum) 
\log\left(\frac{4}{\epsslend^2 \gamma^2} \right)  \dd \thetadum 
-\int_{-\pi}^{\pi}
c(s,\thetadum) \dd \thetadum \nonumber
  \\
&-\crossradius(s)\int_{-1}^{1}\frac{\dd q}{|q|}
\int_{-\pi}^{\pi} \activity(s,\thetadum)   \dd \thetadum 
+	~\int\limits_{-1}^{1}\!\int\limits_{-\pi}^{\pi}\frac{\crossradius(s+q)\activity(s+q,\thetadum)}{ |\bvRo(s,s+q)|}    
 \dd \thetadum \dd q.
\end{align}

Noting again that the log term can be expanded as in Eq.~\eqref{eq:LogExpansion},
we find that the leading-order concentration field satisfies
\begin{align} \label{eq_leadingorderconc}
	2\pi c^{(0)}(s,\theta) + \langle
c^{(0)}(s) \rangle
=&	~\int\limits_{-1}^{1}\! \left[
\frac{\crossradius(s+q)\langle\activity(s+q)\rangle}{ |\bvRo(s,s+q)|}
-\frac{\crossradius(s)\langle\activity(s)\rangle }{|q|}  \right]
 \dd q   
  \\
 &+
\crossradius(s)	\langle\activity(s) \rangle \log 
\left( 
\frac{2}{\epsslend^2  \crossradius^2(s)}\right)  
 -
\crossradius(s)	\int_{-\pi}^{\pi}\activity(s,\thetadum) 
     \log\left[1-\cos(\theta-\thetadum)\right] \dd \thetadum. \nonumber
	\end{align}
where $\langle f(s) \rangle = \int_{\pi}^{\pi} f(s,\theta) \dd \theta$ is the $\theta$-integrated value of a function $f(s,\theta)$.
Integrating this over $\theta$, we find that
\begin{align}
	2\langle\zerothorder{c}(s)\rangle
= & 
 \int_{-1}^{1} \left[
\frac{\crossradius(s+q)\langle\activity(s+q)\rangle}{ |\bvRo(s,s+q)|} 
-\frac{\crossradius(s)\langle\activity(s)\rangle}{|q|} \right]
 \dd q
+
\crossradius(s)\langle\activity(s)\rangle\log\left(\frac{4 }{\epsslend^2 \crossradius^2(s)}\right), 
\end{align}
where we have used that $\int_{-\pi}^{\pi}\log\left[1-\cos(\theta-\thetadum)\right]\dd \thetadum=-2\pi\log(2)$. This can be substituted back into Eq.~\eqref{eq_leadingorderconc} to
obtain the leading-order Slender Phoretic Expression for Looped Active Filaments:
\begin{multline}
	2\pi c^{(0)}(s,\theta) 
=	~\frac{1}{2}\int\limits_{-1}^{1}\! \left[
\frac{\crossradius(s+q)\langle\activity(s+q)\rangle}{ |\bvRo(s,s+q)|}
-\frac{\crossradius(s)\langle\activity(s)\rangle }{|q|}  \right]
 \dd q   -
\crossradius(s)	\langle\activity(s) \rangle \log 
\left( \epsslend  \crossradius(s)\right)   
\\
-
\crossradius(s)	\int_{-\pi}^{\pi}\activity(s,\thetadum) 
     \log\left[1-\cos(\theta-\thetadum)\right] \dd \thetadum.
\end{multline}


Note that, for an axisymmetric activity, $\activity(s,\theta)\equiv\activity(s)$, the $\theta$-integrated activity is simply $\langle\activity(s,\theta) \rangle = 2\pi \activity(s)$. In this case, we therefore find a simplified leading-order expression for the concentration 
\begin{align}
	2 \zerothorder{c}(s,\theta)
= & 
\int_{-1}^{1} \left[
\frac{\crossradius(s+q)\activity(s+q)}{ |\bvRo(s,s+q)|} 
-\frac{\crossradius(s)\activity(s)}{|q|} \right]
 \dd q 
+
\crossradius(s)\activity(s)\log\left(\frac{4 }{\epsslend^2 \crossradius^2(s)}\right).\label{c_zerothorder}
\end{align} 

\subsubsection{Next-order concentration}

Collecting the $O(\epsslend)$ terms from Eq.~\eqref{eq:FullExpansion}, we find that
\begin{align}
	2\pi  c^{(1)}(s,\theta)  +& \langle c^{(1)}(s) \rangle
\nonumber
\\	
=&
\frac{1}{2} \crossradius^2(s) \curvature(s)	\int_{-\pi}^{\pi} 
 \activity(s,\thetadum)
 \Bigg[ 
\cos\dtheta(s,\theta) \log\left(\frac{4}{\epsslend^2 \gamma^2 e^2}\right)  
+\cos\dtheta(s,\thetadum)  \log\left( \frac{\epsslend^2 \gamma^2}{4 e^2}\right)  
\Bigg]
\dd \thetadum
\nonumber
\\ 
&+\frac{1}{2}\crossradius (s)\curvature(s)\int_{-\pi}^{\pi}
c^{(0)}(s,\thetadum)\Bigg\{
  - 
  \cos\dtheta(s,\theta)
+  \cos\dtheta(s,\thetadum) 
    \left[ 1+\log\left(\frac{4}{\epsslend^2 \gamma^2 e^2}\right)       \right]
    \Bigg\}  
    \dd \thetadum
    \nonumber
  \\
& %
   +\frac{1}{2}\crossradius^2(s)\curvature(s)\int_{-1}^{1} \frac{\dd q}{|q|}\int_{-\pi}^{\pi}   \activity(s,\thetadum)  \left[
\cos\dtheta(s,\thetadum)-\cos\dtheta(s,\theta)
\right] \dd \thetadum 
\nonumber
  \\
&-	 \int\limits_{-1}^{1}\!\int\limits_{-\pi}^{\pi}\frac{\crossradius(s+q)\activity(s+q,\thetadum)}{ |\bvRo(s,s+q)|}    
  \left[\crossradius(s+q)\curvature(s+q)\cos\Theta(s+q,\thetadum)+\frac{\bvRo}{|\bvRo|^2}\cdot\Deltaerhorhoq \right]     \dd \thetadum \dd q
\nonumber
\\
&+\int\limits_{-1}^{1}\!\int\limits_{-\pi}^{\pi}\frac{\crossradius(s+q)c^{(0)}(s+q,\thetadum)}{ |\bvRo|^3}
\bvRo\cdot\erho(s+q,\thetadum) 
\dd \thetadum \dd q.
\end{align}

If the activity is axisymmetric, then this expression can be simpified since some of the integrals vanish, for example $\int_{-\pi}^{\pi} \erho(s,\thetadum) \dd \thetadum = 0$ and $\int_{-\pi}^{\pi} \cos\dtheta(s,\thetadum) \dd \thetadum = 0$. In addition, $\gamma$ can be be expanded using Eq.~\eqref{eq:LogExpansion}. The first-order concentration field can then be rewritten as 
  	\begin{align}
	2\pi  c^{(1)}(s,\theta)  +& \int_{-\pi}^{\pi} c^{(1)}(s,\thetadum)\dd \thetadum 
\nonumber
\\
=& 
\frac{1}{2} \crossradius^2(s) \curvature(s) \activity(s)	
 \Bigg\{ 
\cos\dtheta(s,\theta) \bigg[2\pi\log\left(\frac{2}{\epsslend^2 \crossradius^2(s) e^2} \right)  -\int_{-\pi}^{\pi} \log\left[1 - \cos (\theta - \thetadum)\right] \bigg]\dd \thetadum  
\nonumber
\\
&\qquad\qquad\qquad\qquad\qquad\qquad
+
\int_{-\pi}^{\pi}
\cos\dtheta(s,\thetadum)  \log\left[1-\cos(\theta - \thetadum)\right]\dd \thetadum 
\Bigg\}
\nonumber
\\ 
&-\frac{1}{2}\crossradius (s)\curvature(s)c^{(0)}(s) 
\Bigg\{
  2\pi \cos\dtheta(s,\theta)
+ \int_{-\pi}^{\pi} \cos\dtheta(s,\thetadum) 
      \log\left[1 - \cos(\theta - \thetadum) \right]    
      \dd \thetadum 
    \Bigg\}  
\nonumber
  \\
& 
   -\pi\crossradius^2(s)\curvature(s)\activity(s)\cos\dtheta(s,\theta)  \int_{-1}^{1} \frac{\dd q}{|q|}
   \nonumber
  \\
&-	 2\pi\crossradius(s) \int\limits_{-1}^{1}\! \frac{\crossradius(s+q)\activity(s+q)}{ |\bvRo(s,s+q)|^3}    
  \bvRo(s,s+q)   \dd q \cdot\erho(s,\theta).
	\end{align}
 
The integrals in $\thetadum$ can be evaluated to give
\begin{align}
    \int_{-\pi}^{\pi} \log[1-\cos(\theta-\thetadum)] \dd \thetadum =& -2\pi\log2, \\
    \int_{-\pi}^{\pi} \log[1-\cos(\theta-\thetadum)] \cos\dtheta(s,\thetadum) \dd \thetadum =& -2\pi\cos\dtheta(s,\thetadum).
\end{align}
 
Finally, we note that when this expression is integrated over $\theta$ we find that $\int_{-\pi}^{\pi}c^{(1)}(s,\theta) \dd \theta = 0$; hence the $O(\epsslend)$ contribution to the surface concentration for an axisymmetric activity is
\begin{multline}
   c^{(1)}(s,\theta)  
    =
     \frac{1}{2} \crossradius^2(s) \curvature(s) \activity(s)	
    \cos\dtheta(s,\theta)  \left[\log\left(\frac{4}{\epsslend^2 \crossradius^2(s)} \right)
    - 3\right]
\\ 
- \crossradius(s) \int\limits_{-1}^{1}\!
\left[ 
\frac{\crossradius(s+q)\activity(s+q)}{ |\bvRo(s,s+q)|^3}    
      \bvRo(s,s+q)    \cdot\erho(s,\theta)
  +  \frac{\crossradius (s)\curvature(s)\activity(s)\cos\dtheta(s,\theta)}{2|q|}\right]\dd q  .   
\end{multline}

\subsection{Slip velocity}
\label{App_SlipVel}

The normal  to the filament's surface, pointing out of the filament, is
 \begin{align}
     \no(s,\theta) 
&=
     \frac{ 
\left[1- \epsslend\crossradius(s) \curvature(s) \cos\dtheta   \right]\erho(s,\theta ) - \epsslend   \ddrhods \tanhat(s) 
}{\sqrt{ 
\left[1-\epsslend\crossradius(s) \curvature(s) \cos\dtheta \right]^2 + \epsslend^2 \left(\ddrhods \right)^2 }}. \label{normal_filament_full}
 \end{align}
Combining the boundary conditions for the activity and slip velocity
allows us to write the slip velocity as 
 \begin{align}
	\vslip &= \mobility(\bv{x}) \left(\idmat - \no \no \right)\cdot\boldsymbol{\nabla} c 
	= \mobility  \left(\boldsymbol{\nabla} c + \no \activity  \right)  
	\qquad \text{on } S,
	\label{vslip_nBC}
\end{align}
since $\activity = -\no\cdot\nabla c$.
Using the local cylindrical polars (with the axisymmetry axis along $\tanhat(s)$), we then find that the concentration gradient can be written as   
\begin{align}
    \boldsymbol{\nabla} c 
    &= 
    \tanhat (\tanhat \cdot \boldsymbol{\nabla} c )
    +   \etheta (\etheta \cdot \boldsymbol{\nabla} c )
    +  \erho (\erho \cdot \boldsymbol{\nabla} c ) 
   =  \tanhat \partial_s c
    +   \etheta \frac{1}{\epsslend \crossradius}\partial_\theta c  
    +  \erho (\erho \cdot \boldsymbol{\nabla} c ),  \label{nablacEq}
\end{align}
where we have used that $\tanhat\cdot\boldsymbol{\nabla}c = \partial_s c$ and $\etheta\cdot\boldsymbol{\nabla}c = \frac{1}{\epsslend \crossradius(s)} \partial_\theta c$.

For the final term in Eq.~\eqref{nablacEq}, we apply the activity boundary condition, which says that $\no   \cdot \nabla c = -\activity $. This means that $\erho \cdot \boldsymbol{\nabla} c = -\activity + O(\epsslend)$, and so the slip velocity can be rewritten as
\begin{align}
\frac{1}{\mobility}	\vslip 
	&= \tanhat \partial_s c
    +   \etheta \frac{1}{\epsslend \crossradius}\partial_\theta c  
    +  \erho (\erho \cdot \boldsymbol{\nabla} c )
    +
    \no \activity   
    \nonumber\\
    	&= \tanhat \partial_s c
     +   \etheta \frac{1}{\epsslend \crossradius}\partial_\theta c  
     + O(\epsslend).
\end{align}
This gives the required result, as given in the main text.

\section{Glazed torus with sinusoidal activity} \label{sec:GlazedSinusoid}
 
An activity $\activity = -\sin\theta$ decomposed into a Fourier series has only one non-zero coefficient: the first sine mode of the activity, $\Acoeffsinmode{1} = -2\pi$. We therefore find that the leading-order surface concentration is given by 
 \begin{equation}
       \zerothorder{c} = -\sin \theta,
\end{equation}
and the leading-order slip velocity is 
\begin{align}
        \vslip(s,\theta)  
       &= -\frac{\mobility}{  \epsslend  } \etheta(s,\theta) 
    \cos  \theta    +O(1) 
       = -\frac{\mobility}{  2\epsslend  } \left[(1 + \cos 2\theta) \binorhat(s) - \sin 2 \theta \norhat(s)\right]     +O(1),
\end{align}
since $\etheta(s,\theta) = \binorhat(s) \cos\theta  - \norhat(s) \sin\theta $. 
In terms of our Fourier modes, we find that the zeroth mode for the slip velocity is then
\begin{align} \label{vslip_zeromode_glazedsinusoid}
       \vslipzeromode(s) &=  -\frac{\pi\mobility}{   \epsslend  } \bv{e}_z.
\end{align}
Using the same arguments as for the Janus glazed torus, the swimming speed must exactly compensate for the (uniform) slip velocity, and so we determine the swimming speed of the sinusoidal glazed torus (with uniform mobility) to be
\begin{align}
    \Uswim = \frac{ \mobility}{   2\epsslend  } \bv{e}_z.
\end{align}

\section{Comparison with Boundary Element Methods}

\begin{figure}
    \centering
    \subfloat[Uniform torus concentration]{\includegraphics[width=0.45\textwidth,viewport=50 550 300 750,clip]{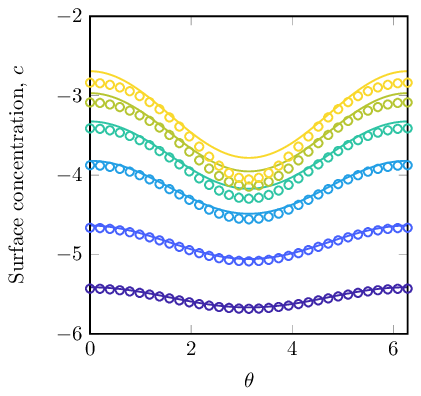}}
    \subfloat[Uniform torus concentration error]{\includegraphics[width=0.45\textwidth,viewport=50 550 300 750,clip]{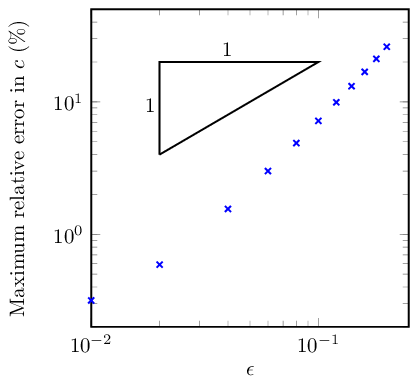}}
    
    \subfloat[Glazed torus concentration]{\includegraphics[width=0.45\textwidth,viewport=50 550 300 750,clip]{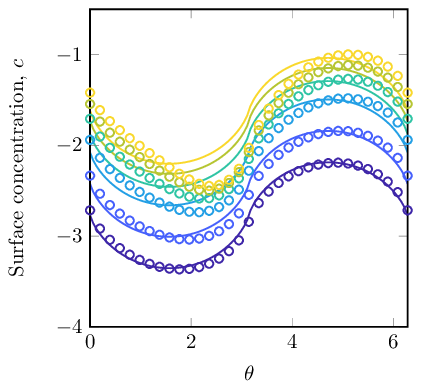}}
    \subfloat[Glazed torus concentration error]{\includegraphics[width=0.45\textwidth,viewport=50 550 300 750,clip]{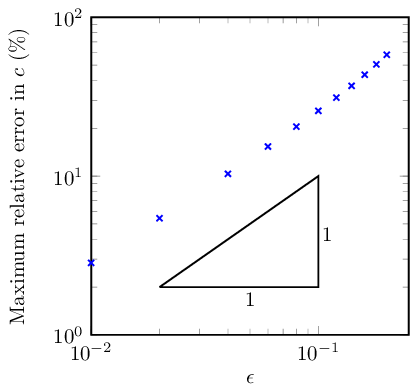}}
    
    \caption{Comparison between boundary element solutions (circles) and analytic solutions to SPT (curves) for slenderness $\epsslend=0.01,0.02,0.04,0.06,0.08,0.1$ (dark blue to yellow). 
    (a) The surface concentration for a uniform torus as a function of $\theta$, compared to the analytic result, Eq.~\eqref{eq:UniformDonut_Conc}. 
    (b) The percentage relative error between analytic and BEM solutions decays faster than $O(\epsslend)$.
    (c) The surface concentration for a glazed torus as a function of $\theta$, compared to the analytic result, Eq.~\eqref{eq:GlazedTorusConc}.
    (d) The percentage relative error between analytic and BEM solutions decays like $O(\epsslend)$.
    }
    \label{fig:BEMcomparison}
\end{figure}

To validate the results of our Slender Phoretic Theory for Looped Filaments, we compare our analytic solutions to those generated using a boundary element method (BEM).

In Fig.~\ref{fig:BEMcomparison}, we show the BEM results for a uniformly active torus and a glazed torus for a range of values of the slenderness, compared to the analytical solutions given in Eqs.~\eqref{eq:UniformDonut_Conc}~\&~\eqref{eq:GlazedTorusConc}. We also plot the maximum percentage difference between the two, and find that the error decays superlinearly with $\epsslend$ for the uniform torus and linearly for the glazed torus. These are exactly as expected, since the analytical expressions given are valid up to $O(\epsslend)$ in the uniform case, and $O(1)$ (just the leading order term) in the glazed case.

Note that we also calculate the swimming velocity for the glazed torus using BEM, and find that in the range $\epsslend\in[0.02,0.2]$ the calculated swimming speed differs from the analytical solution \eqref{eq:JanusGlazedSpeed}, $U=1/\pi\epsslend$, by less than $5\%$.

\bibliography{main}

\end{document}